# Bottleneck Congestion And Work Starting Time Distribution Considering Household Travels

Qida SU, David Z.W. WANG[*]

School of Civil and Environmental Engineering, Nanyang Technological University,
50 Nanyang Avenue, Singapore 639798

**Abstract**

Flextime is one of the efficient approaches in travel demand management to reduce peak hour congestion and encourage social distancing in epidemic prevention. Previous literature has developed bi-level models of the work starting time choice considering both labor output and urban mobility. Yet, most analytical studies assume the single trip purpose in peak hours (to work) only and do not consider the household travels (daycare drop-off/pick-up). In fact, as one of the main reasons to adopt flextime, household travel plays an influential role in travelers' decision making on work schedule selection. On this account, we incorporate the household travels into the work starting time choice model in this study. Both short-run travel behaviours and long-run work start time selection of heterogenous commuters are examined under agglomeration economies. If flextime is not flexible enough, commuters tend to agglomerate in work schedule choice at long-run equilibrium. Further, we analyze optimal schedule choices with two system performance indicators. For total commuting cost, it is found that the rigid school schedule for households may impede the benefits of flextime in commuting cost saving. In terms of total net benefit, while work schedule agglomeration of all commuters leads to the maximum in some cases, the polarized agglomeration of the two heterogenous groups can never achieve the optimum.

[*] Corresponding author. Tel: (65) 6790-5281

Email: wangzhiwei@ntu.edu.sg   qida.su@ntu.edu.sg

1. **Introduction**

Peak hour traffic congestion in metropolitan areas has always been a headache to both city governors and travelers. The emergence of peak hours during workday is largely credited to the traditional rigid work schedules, in which commuters need to arrive at workplace with a rigid work starting time and can leave for home after a rigid work quitting time. From travel demand management's perspective, flextime/staggered work hours (SWH) as a flexible work arrangement (FWA) could intuitively be an effective approach to decentralize the travel demand and thereby alleviate traffic congestion. In terms of the epidemic prevention, flextime/SWH is also to the benefits of the social distancing in transportation facilities and working areas. Previous literature has shown other advantages of flextime/SWH (Ralston, 1989; Christensen & Staines, 1990; Ezra & Deckman, 1996; Lucas & Heady, 2002; Gainey & Clenney, 2006), e.g., increase in employees' retention.

Highly recognized in literature, the flextime/SWH are still not widely adopted by employers and employees, especially in the Asia Pacific. Take Singapore as an example, according to the report in Ministry of Manpower (MOM, 2020), only 29.8% of firms in both public and private sectors offer flextime/staggered hours in 2019.[1] One crucial obstacle preventing the promotion of flextime/SWH is the concern of employers about productivity. Regarding economic agglomeration, such flextime/SWH staggers the working hours of employees, and hence the less agglomerated work hours lead to the lower instantaneous productivity of each employee. As for employees, whether to choose a firm offering flextime/SWH depends on numerous factors. A survey carried out by Karin and Danielle (2007) showed that in Asia Pacific area the most chosen reason to adopt flextime is to avoid congestion (89%), followed by household travels (daycare pick-up/drop off) (57%).

To better investigate people's work schedule choice and the corresponding travel behavior, several analytical models have been developed in literature. Seminal work is carried out by Henderson (1981). Incorporating the productivity and traffic congestion effect, he examined the equilibrium and optimal scheme of staggered work hours. He showed that work starting times are distributed continuously at equilibrium. Following Henderson's framework, Wilson (1992) scrutinized the choice of residential location,

---

[1] 52.7% firms offer formal FWA include part-time work, flextime/staggered hours, formal tele-commuting, homeworking, job sharing, compressed work week.

Arnott et al. (2005) generalized the model to include firms' heterogeneity and Arnott (2007) explored the potential of congestion tolling in congestion alleviation. Later Mun and Yonekawa (2006) focused on the flextime where commuters can choose their work starting time freely. He modeled the traffic congestion via Vickrey (1969)'s bottleneck model and tried to figure out the equilibrium and an optimum number of firms using flextime. Takayama (2015) reformulated the discrete staggered work hours model with peak hour congestion as a potential game. Fosgerau and Small (2017) studied commuters' trip timing preference incorporating the productivity effects of work and leisure. Moreover, Zhu et al. (2018) proposed an activity-based bottleneck model to analyze the optimal work starting time. More recently, Su and Wang (2020) focused on the compressed work schedule to examine its effects on travelers' utility in a bi-level model.

However, one assumption usually made in previous studies is that all trips during peak hour share same single purpose only traveling from home to workplace and hence traffic congestion was caused solely by work trips in their models. Even for those encapsulating activity utilities (Li et al., 2014; Fosgerau et al., 2017; Zhu et al., 2018), they merely examined the respective utility at home and workplace, i.e., the utility at origin and destination. To the best of our knowledge, trips with other purposes (e.g., household travels) were seldom addressed. Their effects on traffic congestion and work schedule choices remain unknown.

In fact, as one of the main reasons to adopt flextime/SWH (Karin et al., 2007), the household travels, which generally consist of multiple trips among family members: the daycare pick-up/drop-off and the work commute, are very common and can lead to significant peak hour congestion, in particular in cities with high car ownership threshold, e.g., Singapore. Furthermore, some countries such as U.S and Singapore(CDC, 2020; ECDA, 2020) also advice parents to pick-up/drop-off their daycare children amid the pandemic.

Without considering household travels, the merits of flextime/SWH in congestion alleviation could often be overestimated. Recently incorporating the household travels, Jia et al. (2016) investigated the dynamic equilibrium travel behavior. Liu et al. (2016) extended the study to the case with mixed travelers and proposed the optimal schedule of school and work to reduce traffic congestion. Moreover, Zhang et al. (2017) revisited the household travels with a different school location choice. However, the work schedule choices were scarcely discussed in these studies. Only in Zhang et al. (2017)

the differentiated school and work schedules were examined preliminarily without the consideration of productivity effect.

Specifically, this paper aims to examine both household and individual travels in the interrelationship between bottleneck congestion and work starting time choice behavior. Following Mun et al. (2006), we assume that commuters can choose their work starting time freely within a specified period (*flex-interval*) if flextime/SWH are offered. A bi-level economic model is developed enclosing positive production externalities and negative congestion externalities with commuters of multiple trip purposes. Both the short-term equilibrium with within-day travel behavior, and the long-term equilibrium with work schedule choices are analyzed. The question on how household travelers differ from the individual travelers in work schedule choice is being addressed. We further inspect the effects of different flextime design on different travelers' work schedule choice and determine the system optimum in both total commuting cost minimization and total net benefit maximization.

The rest of the paper is organized as follows. Section 2 presents the model framework incorporating the production and traffic congestion effect of mixed travelers. The short-term equilibrium is examined in Section 3, followed by the long-term equilibrium discussed in Section 4. Next, the optimal compositions of flextime users are studied in Section 5. Numerical examples are also provided in Section 6. Finally, Section 7 wraps up the paper with some concluding remarks and future research directions.

2. **Model formulation**

*2.1 Basic setting*

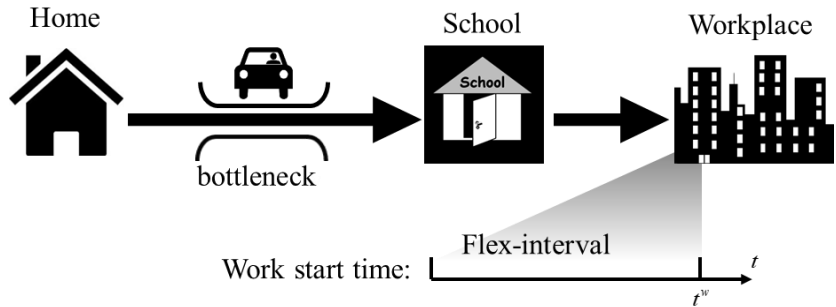

*Figure 1 A monocentric city with home-bottleneck-school-workplace network.*

The list of notations throughout the process of model formulation is presented in Appendix A. Consider a city with a CBD and a residential area connected by a highway with a single bottleneck, and there is a school in between after the bottleneck (Figure 1). A fixed total number, $N$, of roadway travelers commute between home and workplace in CBD every day.

Two heterogenous groups of commuters are considered, wherein $\mu \in [0,1]$ of them are household (H) and the rest are individual (I), and the following assumptions are made.

***A1.*** *Individual commuters $N_I = (1-\mu)N$ travel to workplace directly with a desired arrival time/interval depending on one's work schedule.*

***A2.*** *In addition to the trip to work, household commuters $N_H = \mu N$ need to send and pick up their children to/from school with fixed school starting time $t^s$ (Jia et al., 2016; Liu et al., 2016). No virtual or remote learning is considered.*

***A3.*** *Each vehicle takes only one commuter and one child at most.*

Here, A1 and A2 distinguishes the commuter groups with heterogeneity in trip purpose and A3 ensures that the vehicles demand equals commuters demand with $N = N_H + N_I$.

As for commuters' work schedule, we examine two schedule options: normal (rigid) work hours (r) and flextime (f), to focus on the temporal effects of work schedule choices with the following assumptions.

***A4.*** *$N^r$ employees in normal working hours start work at a fixed time point $t^w$ ($t^w > t^s$) and work $H$ $\left(H > 2\dfrac{N}{s}\right)$ hr/day[2].*

---

[2] The daily work hour $H$ is assumed to be longer than the total time of both morning and evening commuting periods.

**A5.** $N^f \left(= N - N^r\right)$ employees in flextime also work $H$ hr/day with uniform flex-interval $\Delta \left(\leq \frac{N}{s}\right)$[3] in which they can choose to start work freely before the last work starting time (Ben-Akiva et al., 1984; Xiao et al., 2014). They can start work immediately once reach the workplace within the flex-interval while early arrivals need to wait till the start of flex-interval.

**A6.** *No late arrival is allowed for all employees and the latest work starting time for flextime employees coincides with $t^w$.*

To note, A4 and A5 describe the schedule design differences of the two schedule and hence employees in flextime can enjoy flexibility in work starting time selection. The relative length of *flex-interval* is defined as $\theta = \frac{\Delta}{N/s}$. Evidently, $\theta \in (0,1]$ and $\theta = 1$ when $\Delta$ reaches its maximum. Besides, A6 ensures all employees must be at work at $t^w$ to better evaluate their productivity. From A6, the core hours during which all employees must be at work can be calculated as $H - \Delta$ and the *flex-interval* starts at $t^w - \Delta$.

Further, we let $v_H, v_I \in [0,1]$ denote the proportion of normal employees among household and individual travelers, respectively. Hence, the number of employees of each group in each work schedule can be derived readily, where

$$N_H^r = v_H \mu N, \quad N_H^f = (1-v_H)\mu N, \quad N_I^r = v_I(1-\mu)N, \quad N_I^f = (1-v_I)(1-\mu)N. \quad (1)$$

### 2.1.1 Employers and employees' behavior

Now we formulate the work starting time choice in a bi-level model. In the lower level, the employees(commuters) can change neither their job nor work schedule and they seek to maximize their utility by changing their trip timings of commuting and schooling (for household travelers). The short-run equilibrium is reached when no employees in the same work schedule with same trip purposes can reduce their aggregate travel cost $P$ unilaterally (See Section 3). In the upper level, employers and employees can decide whether to adopt or work in normal work hours or flextime to maximize their perceived benefit. That is to say, $\{v_H, v_I\}$ are endogenously determined by the long-run choice behaviors (See Section 4).

---
[3] While it is possible to have $\Delta > \frac{N}{s}$, such that the *flex-interval* is long enough and there is no congestion cost for flextime employees even if all commuters choose flextime, we bounds the longest *flex-interval* to focus on the scenarios with bottleneck congestion.

From employers' perspective, whether to adopt flextime depends on net profit per employee, which is defined as the difference between the employees' productivity and the corresponding wage expenses,

$$\pi^i = F^i - w^i, \quad (2)$$

and all firms aim to maximize it. On the other hand, from employees' perspective, whether to work in a firm adopting flextime depends on the net income, which is the difference between wage income and equilibrium commuting cost,

$$u_J^i = w^i - P_J^i, \quad (3)$$

and each employee seeks to maximize it.

Further, it is assumed that firms produce homogeneous goods under perfect competition and employees with identical skills can switch firms freely, which suggests that all firms are price takers at long-run equilibrium. Under perfect competition, all of them earn zero profit and thus have no incentive to alter their work schedule choice, where,

$$\pi^r = \pi^f = 0, \rightarrow F^r = w^r, F^f = w^f, \quad (4)$$

$$u_H^r = u_H^f, u_I^r = u_I^f. \quad (5)$$

Therefore, for group $J$ commuters, in the case when both schedules are used at equilibrium, the individual net benefits of that type of employees are equal, regardless of the employer type (normal schedule or flextime), i.e.,

$$\delta_J^r = F^r - P_J^r = F^f - P_J^f = \delta_J^f. \quad (6)$$

And in the case when there is only one schedule $i^*$ adopted,

$$\delta_J^{i=i^*} = F^{i=i^*} - P_J^{i=i^*} > F^{i\neq i^*} - P_J^{i\neq i^*} = \delta_J^{i\neq i^*}. \quad (7)$$

Note that the model enclosing both employees' and employers' behaviors is similar to those in (Mun et al., 2006; Takayama, 2015; Su et al., 2020). Before we scrutinize the properties of the bi-level equilibrium in Section 3 and 4, we first formulate the productivity and commuting cost of employees in Section 2.2 and 2.3 .

*2.2 Productivity*

Firms agglomerate spatially and temporally in the CBD areas to boost the agglomeration economies. One typical assumption in the literature on temporal agglomeration is that the unit time output per employee is independent of daily working hours and increases with the number of employees on duty at that time unit (Henderson, 1981; Takayama, 2015). Namely, in terms of productivity, it is assumed that there is positive externality to agglomerate work hours of employees. Following this, for normal schedule employees, their individual instantaneous output at time $t$ is defined as an increasing function of the total number of employees on duty at time $t$, $N^r(t) + N^f(t)$, and is given as

$$y(t) = \kappa \cdot f\left(N^r(t) + N^f(t)\right), \tag{8}$$

where $\kappa > 0$ is a constant parameter and $f(\cdot)$ is a shift factor representing agglomeration economy with $f'(\cdot) > 0$. Note that when all commuters are on duty, the instantaneous output $y(t)$ reaches maximum equal to $\kappa f(N)$. The individual outputs per day for employees with normal schedule are then given as follows.

$$F^r = \int_{t^w}^{t^w + H} y(t)\, \mathrm{d}t = \kappa \int_{t^w}^{t^w + H} f\left(N^r(t) + N^f(t)\right) \mathrm{d}t. \tag{9}$$

As for flextime employees, their individual instantaneous output at time $t$ is defined in the same form as in Eq.(8)[4]. Admittedly, the daily individual outputs of different flextime users differ depending on one's work starting time. However, from employer's perspective, it is the total productivity of all flextime employees that matters, and undifferentiated wages are paid to all. Given the undifferentiated wages, the net benefit perceived by the same group users in flextime remain unchanged, regardless of one's actual output. On this account, we take the average as the individual output of flextime employees, as follows.

---
[4] In fact, there exists intra-firm productivity effect, where the output of firms adopting flextime may be lower when some employees are not present during *flex-interval*s. For simplicity, we assume that the intra-firm productivity effect roots in the total labor input only, i.e., the number of total employees on duty, and hence the individual instantaneous output is the same as that in normal schedule.

$$F^f = \frac{1}{N^f} \left( N^{fe} \int_{t^w-\Delta}^{t^w-\Delta+H} y(t)\, \mathrm{d}t + \sum \int_{t'}^{t''} \left( N^{f\,\prime}(\omega) \int_{\omega}^{\omega+H} y(t)\, \mathrm{d}t \right) \mathrm{d}\omega \right), \qquad (10)$$

where $N^{fe}$ represents the number of flextime commuters who arrive at workplace early before the start of *flex-interval*, and $t', t''$ denote the first and last work starting time of a continuous period, in which flextime employees start work continuously within the *flex-interval*. Moreover, $N^{f\,\prime}(t)$ corresponds to the number of flextime employees who start to work at time $t$, and thus $N^{f\,\prime}(\omega) \int_{\omega}^{\omega+H} y(t)\, \mathrm{d}t$ is the total daily output of commuters starting work at $\omega$. To note, $N^f(t'') - N^f(t')$ is the number of flextime commuters starting work from $t'$ to $t''$. Multiple $t'$ and $t''$ may exist depending on the actual arrival patterns at workplace.

*2.3 Commuting cost*

The daily departure pattern is determined by work schedules as well as the school schedule. To better investigate the dynamics of congestion effect resulting from the two work schedules on the two groups of commuters, we apply Vickrey (1969)'s bottleneck model, where the bottleneck with capacity $s$ follows FIFO principle. The CBD area is abstracted as a point with all firms located at that point and the school is abstracted as a point right after the bottleneck. Without loss of generality, free-flow travel time between home and school, and between school and workplace are assumed to be zero (Liu et al., 2016).

All employees are assumed to share the same value of time $\alpha \left( \leq \kappa f(N) \right)$[5] and the same scheduling penalty for a unit time of early arrival $\beta$, which satisfy $\alpha > \beta > 0$ (Small, 1982). We further denote $\rho = \dfrac{\beta}{\kappa f(N)}$ to represent the relative schedule penalty compared to the instantaneous individual output $(0 < \rho < 1)$. Over the course of morning peak, individual commuters determine his/her home departure time depending on the queuing cost and the scheduling penalty cost at workplace. As for household travelers, they make a joint decision choice on their home departure time considering both work and school trips with preferred arrival time $t^s$ and $t^w$. Following (Jia et al., 2016; Liu et al., 2016), the value of time and the unit time scheduling penalty for the child in household travels are simply assumed to the same as those for employees. Furthermore, the commuting cost for household travelers is calculated as the composite travel costs of both the child's school trip and parent's work trip.

---

[5] From definition, the value of time $\alpha$ is the monetary amount one employee would be willing to pay to reduce travel time and can be considered as unit time income. In our model, as the individual wage equals to individual output at equilibrium and the maximum unit time output is $\kappa f(N)$, there is $\alpha \leq \kappa f(N)$, where the equality sign holds if and only if no employee adopts flextime.

Therefore, the commuting costs for individual and household employees in normal schedule departing at time $t$ are

$$p_I^r(t) = \alpha T(t) + \beta \max\{0, t^w - t - T(t)\}, \quad (11)$$

$$\begin{aligned} p_H^r(t) &= \alpha T(t) + \beta \max\{0, t^s - t - T(t)\} \\ &+ \alpha T(t) + \beta \max\{0, t^w - t - T(t)\}. \end{aligned} \quad (12)$$

Similarly, the commuting costs for individual and household employees in firms adopting flextime departing at time $t$ are

$$p_I^f(t) = \alpha T(t) + \beta \max\{0, (t^w - \Delta) - t - T(t)\}, \quad (13)$$

$$\begin{aligned} p_H^f(t) &= \alpha T(t) + \beta \max\{0, t^s - t - T(t)\} \\ &+ \alpha T(t) + \beta \max\{0, (t^w - \Delta) - t - T(t)\}. \end{aligned} \quad (14)$$

To note, $T(t) = \dfrac{D(t)}{s}$ is the travel time for departure time $t$ on the highway and $D(t)$ is the queue length at the bottleneck at $t$ and should satisfy

$$\frac{\partial D(t)}{\partial t} = \begin{cases} \sum_{\substack{i=r,f \\ J=H,I}} r_J^i(u) - s, & \sum_{\substack{i=r,f \\ J=H,I}} r_J^i(u) > s \text{ or } D(t) > 0; \\ 0, & \sum_{\substack{i=r,f \\ J=H,I}} r_J^i(u) \leq s \text{ or } D(t) = 0, \end{cases} \quad (15)$$

where $r_J^i(t)$ is the flow rate of group $J$ commuters in work schedule $i$ entering the bottleneck at time $t$.

When considering household travels, schedule coordination has been discussed in the literature, and it is found that proper coordination between work and school schedule can achieve minimization of total commuting cost (Liu et al., 2016; Zhang et al., 2017). Given that our focus is on the attractiveness of flextime to household and individual travelers, we let the gap between school and work starting time of normal schedule employees is fixed and given as,

$$t^w - t^s = \frac{N_I}{s}. \quad (16)$$

For comparisons in following chapters, we further consider the case when no commuter adopts flextime $(\mu=0)$ as the benchmark case. One may notice that the school-work schedule gap in Eq.(16) is indeed the optimal gap in total travel cost minimization for the benchmark case, which can be readily obtained from Eq.(13) in (Liu et al., 2016) by setting the unit time late arrival penalty $\gamma \to +\infty$. Any further reduction in total travel cost should be attributed to the introduction of flextime solely.

Note that the preceding discussion focuses on commuting cost during morning peak. For evening peak, following previous studies (Vickrey, 1973; Hurdle, 1981; Hurdle et al., 1983; de Palma & Lindsey, 2002), it is assumed that one's travel cost in evening peak equals to his/her travel cost in the morning. In fact, similar to morning peak, household travelers who need to pick up their children before going back home in the evening, also bear additional cost related to their children (e.g., travel time cost, late pick-up penalties), whereas the individual travelers need not to do so. Besides, it applies to both morning and evening peak that the travel cost of flextime employees is generally smaller than the travel cost of normal schedule employees, as the former can choose their work starting/leaving time freely to avoid the peak hours. On this account, in what follows the daily commuting cost is simply calculated as two times of the cost in morning peak.

## 3. Short-run equilibrium

In this section, the equilibrium travel pattern is analyzed. In short run, employees cannot change their firms so that the numbers of individual and household employees in normal schedule, i.e., $v_H, v_I$ are given and hence $N_H^r, N_H^f, N_I^r, N_I^f$ can be derived. Commuters seek to minimize their travel cost only and at equilibrium the commuting cost of the same group in same work schedule equals.

Derived from within-day travel equilibrium $p'(t) = 0$ and Eqs.(11)-(15), the equilibrium arrival rates at the bottleneck for individual and household commuters with normal schedule are equal and given by

$$r_I^r = r_H^r = \frac{\alpha}{\alpha - \beta} s > s. \tag{17}$$

As for flextime, the equilibrium arrival rates at the bottleneck for individual commuters who arrive workplace before and within the *flex-interval*, respectively, are given by

$$r_I^{fe} = r_I^r; \quad r_I^{fp} = s. \tag{18}$$

Similarly, the equilibrium arrival rates at the bottleneck for household commuters who arrive workplace before and within the *flex-interval* are

$$r_H^{fe} = r_H^r; \quad r_H^{fp} = \frac{2\alpha}{2\alpha - \beta} s. \tag{19}$$

One may refer to (Arnott et al., 1990; Liu et al., 2016) for details derivation of Eqs.(17)-(19). Apparently, $r_I^r = r_H^r = r_I^{fe} = r_H^{fe} > r_H^{fp} > r_I^{fp} = s$. With the above formulations, the four possible dynamic commute patterns at user equilibrium are derived and shown in Figure 2. The occurrence conditions of each pattern are summarized in Table 1.

While Figure 2 delineates the arrivals at and departures from the bottleneck (solid and dashed black curves), it also depicts the equilibrium work starting time before $t^w$ (red solid curves) of travelers in both work schedules. As zero free flow travel time between school and workplace is assumed, the departure time from bottleneck coincides with the arrival time at school and workplace, which also coincides the work start time for flextime employees if they arrive at workplace within the *flex-interval*. Nonetheless, the work start time for normal employees remains unchanged as $t^w$.

From Figure 2, it should also be noted that the two clusters of travels (individual and household) are totally separated. The bottleneck can digest all the individual flows within the assumed schedule gap $t^w - t^s$ and thus individual travelers will depart only after $t^s$. We next distinguish four cases of equilibrium commute patterns as well as the corresponding work start time of flextime employees according to the value of $v_H, v_I$ and $\theta$ (or $\Delta$).

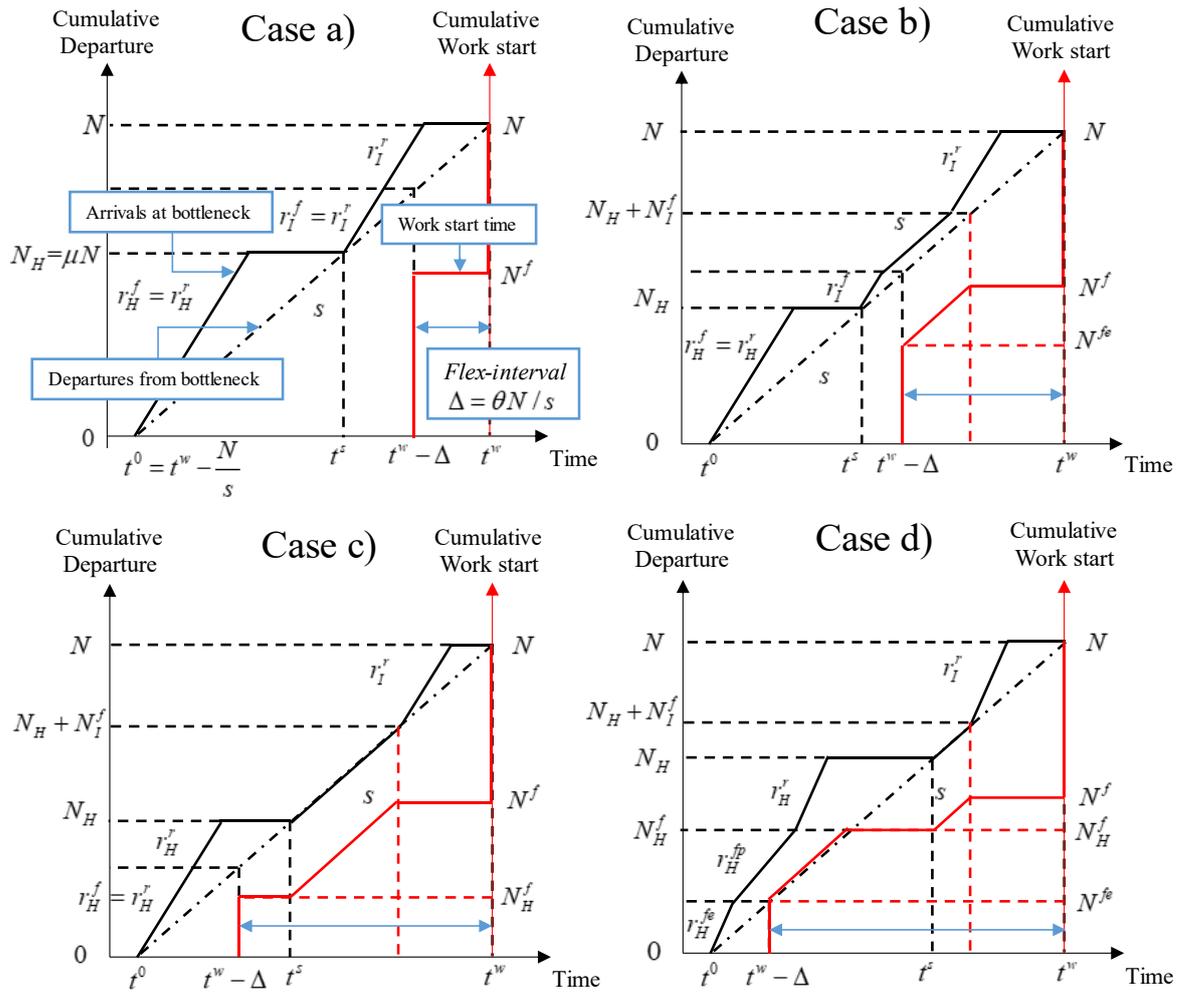

*Figure 2 Possible equilibrium commute patterns and work starting time patterns.*

**Case a):** In this case, there is $\Delta \in \left(0, \frac{N_I^r}{s}\right]$ which is equivalent to $\theta + \mu < 1$ and $v_I \geq \frac{\theta}{1-\mu}$ from definitions. All household travelers $N_H$ arrive at school and work early, and all individual flextime travelers $N_I^f$ arrive at workplace before the *flex-interval*. As for individual normal travelers, some of them arrive at workplace before the *flex-interval*. It is worth mentioning that the arrival orders of travelers before the start of *flex-interval* are indeterminate. Namely, during the period from $t^0$ to $t^s$ and from $t^s$ to $t^w - \Delta$, the exact departure rates of different types of travelers remain unknown. However, as long as the total arrival rate at bottleneck equals to $\frac{\alpha}{\alpha - \beta} s$, the overall traffic pattern stays the same, and hence the equilibrium travel cost is determinate. In fact, the commute pattern in Case a) is identical to the case when all travelers adopt normal schedule $(v_H = 1, v_I = 1)$. As for work start time, the indeterminacy in the arrival order will not affect the work start time distribution because all flextime employees arrive workplace before the *flex-interval* and start work at $t^w - \Delta$.

**Case b):** There is $\Delta \in \left(\frac{N_I^r}{s}, \frac{N_I}{s}\right)$ which is equivalent to $\theta + \mu < 1$ and $v_I < \frac{\theta}{1-\mu}$.

All household travelers $N_H$ arrive at school and work early while some individual flextime travelers arrive at workplace in the *flex-interval*. All individual normal travelers arrive at workplace later than the flextime travelers. In other words, the arrival order is determinate. As for work start time, all household employees and some individual flextime employees $N_H^f + N_I^{fe}$ start work at $t^w - \Delta$ while some individual flextime employees $N_I^{fp}$ arrive workplace within the *flex-interval* and start work at a constant rate $s$.

The critical case between Case a) and b) occurs when $v_I = \frac{\theta}{1-\mu}$, where last individual flextime travelers arrive at workplace right at the start of *flex-interval*.

**Case c):** There is $\Delta \in \left( \dfrac{N_I}{s}, \dfrac{N - N_H^f}{s} \right]$ which is equivalent to $\theta + \mu > 1$ and $v_H \geq \dfrac{\theta + \mu - 1}{\mu}$. All individual travelers $N_I$ arrive at bottleneck within the *flex-interval*, and all household flextime travelers $N_H^f$ arrive at workplace before the *flex-interval*. Likewise, though the arrival order before the *flex-interval* of household travelers is indeterminate, the overall commute pattern will still be the same. For work start time, all household flextime employees $N_H^f$ start work at $t^w - \Delta$ and all individual flextime employees $N_I^f$ start work at the constant rate $s$.

Note that from Eq. (16), when $\theta + \mu = 1$, $t^w - t^s = \dfrac{N_I}{s} = \Delta$. The *flex-interval* start right at the school starting time and the commute pattern is the same as in Case c).

**Case d):** There is $\Delta \in \left( \dfrac{N - N_H^f}{s}, \dfrac{N}{s} \right]$ which is equivalent to $\theta + \mu > 1$ and $v_H < \dfrac{\theta + \mu - 1}{\mu}$. All individual travelers $N_I$ as well as the household normal travelers $N_H^r$ arrive at bottleneck within the *flex-interval* while some household flextime travelers $N_H^f$ arrive at workplace before the *flex-interval*. The arrival order is determinate and all household normal travelers arrive at school and workplace later than flextime travelers. For work start time, some household flextime employees start work at $t^w - \Delta$, followed by the rest household flextime employees starting work at the constant rate $s$. After a window period without any employees starting work (during which the household normal employees arrive at workplace), the individual flextime employees start work with the constant rate $s$.

When $v_H = \dfrac{\theta + \mu - 1}{\mu}$, it is the critical case between Case c) and d), where last household flextime traveler departs from the bottleneck right at the start of *flex-interval*.

One should also notice that the queue has dissipated in case c) and d) during the departures of flextime individuals. As $t^w - t^s$ is predetermined by the total number of individual travelers, the flextime individuals can arrive at workplace within the *flex-interval* without incurring any queuing cost and their minimum arrival rate is $s$ [6].

With Eqs. (11)-(14) and the discussion above, the equilibrium commuting cost for each case is then summarized in Table 2 including schedule penalty and queuing delay costs, with respect to $v_H, v_I, \mu, \theta, \beta, N$ and $s$. Detailed derivations of Table 2 is provided in Appendix B.

As shown in Table 2, there is $P_H^r > P_H^f$ in Case a),b) and c) and $P_I^r > P_I^f$ in Case a),c) and d) (with $\mu < 1$ and $\theta > 0$ ). Since $0 \leq v_I < \frac{\theta}{1-\mu}$ in Case b) and $0 \leq v_H < \frac{\theta+\mu-1}{\mu}$ in Case d), one can also easily verify that $P_I^r > P_I^f$ in Case b) and $P_H^r > P_H^f$ in Case d). The equilibrium commuting cost of normal travelers is generally higher than flextime travelers as intuited for both households and individuals.

*Table 1    Conditions for the occurrence of each pattern depicted in Figure 2.*

| Equilibrium | Conditions | | $N^{fe}$ |
|---|---|---|---|
| Case a) | $\theta + \mu < 1$ | $v_I \geq \frac{\theta}{1-\mu}$ | $(1-v_I+(v_I-v_H)\mu)N$ |
| Case b) | | $v_I < \frac{\theta}{1-\mu}$ | $(1-\theta-v_H\mu)N$ |
| Case c) | $\theta + \mu = 1$ | | $\mu(1-v_H)N$ |
| | $\theta + \mu > 1$ | $v_H \geq \frac{\theta+\mu-1}{\mu}$ | |
| Case d) | | $v_H < \frac{\theta+\mu-1}{\mu}$ | $(1-\theta)N$ |

---

[6] If the arrival rate is lower than $s$, the bottleneck cannot service all the individual commuters $N_I$ within the time gap between $t^s$ and $t^w$, the individual commuters left have to depart earlier before $t^s$ traveling together with household commuters, which increase their travel cost. They will choose to travel after $t^s$ at equilibrium, until the arrival rate becomes $s$.

*Table 2  Equilibrium commuting cost of different types of travelers in patterns depicted in Figure 2.*

| Traveler type | Case/Work Schedule | a) | b) | c) | d) |
|---|---|---|---|---|---|
| H | r | $P_H^r = 2(1+\mu)\frac{\beta N}{s}$ | | | $P_H^r = 2(3-\mu-2\theta+2\mu\nu_H)\frac{\beta N}{s}$ |
| H | f | $P_H^f = 2(1+\mu-\theta)\frac{\beta N}{s}$ | | | |
| I | r | $P_I^r = 2(1-\mu)\frac{\beta N}{s}$ | $P_I^r = 2\big((1-\mu)(1+\nu_I)-\theta\big)\frac{\beta N}{s}$ | | $P_I^r = 2(1-\mu)\nu_I\frac{\beta N}{s}$ |
| I | f | $P_I^f = 2(1-\mu-\theta)\frac{\beta N}{s}$ | | | $P_I^f = 0$ |

### 4. Long-run equilibrium

Now we turn to the long-run equilibrium, where the employers can alter their offered work schedule and employees can select the corresponding employers freely to maximize their net benefit. We will first elaborate on the equilibrium conditions in Section 4.1 and focus on various solutions in Section 4.2 -4.4 .

*4.1 Equilibrium conditions*

As discussed in Section 2.1.1, each commuter in the group $J$ (household or individual) cannot increase his/her net benefit $\delta_J^i = F^i - P_J^i$ by changing the type of firms with work schedule $i$ unilaterally at equilibrium. While we have identified the four commute patterns and the commuting cost $P$ at short-run equilibrium, it is necessary to understand the resulting output $F$ in each pattern.

Hereinafter, we specify the shift factor of productivity effects in Eq.(8) as $f(x)=x$. Based on Eqs (8)-(10) and the work start time distribution depicted in Figure 2, the daily individual output can be derived as summarized in Table 3. Detailed derivation can be found in Appendix C. From Table 3, it can be verified that the individual output of employee with the normal schedule is not always larger than that with flextime depending on the value of $\nu_H, \nu_I$.

*Table 3   Daily individual output of different types of travelers in patterns depicted in Figure 2.*

| Case | Normal Schedule ($F^r$) | Flextime ($F^f$) |
|---|---|---|
| a) | $\kappa HN + (\varepsilon\theta)\dfrac{\kappa N^2}{s}$ | $\kappa HN + (-(\varepsilon+1)\theta)\dfrac{\kappa N^2}{s}$ |
| b) | $\kappa HN + \dfrac{1}{2}\left(\theta(\theta+2\mu v_H-2)+(1-\mu)^2 v_I^2\right)\dfrac{\kappa N^2}{s}$ | $\kappa HN + \dfrac{1}{6\varepsilon}\begin{pmatrix}-4\theta^3+(1-\mu)^2 v_I^2(6-9\mu v_H-5v_I+5\mu v_I)\\-6\theta(\mu v_H-1)(\mu(v_H+v_I)-v_I)\\+3\theta^2(2+v_I-\mu(3v_H+v_I))\end{pmatrix}\dfrac{\kappa N^2}{s}$ |
| c) | $\kappa HN + \dfrac{1}{2}\begin{pmatrix}\mu(2\theta(v_H-1)+(\mu-2)(v_I^2-1))\\+v_I^2-1\end{pmatrix}\dfrac{\kappa N^2}{s}$ | $\kappa HN + \dfrac{1}{6\varepsilon}\begin{pmatrix}\mu(-2(6+\mu(4\mu-9))-6\theta(2+\mu(v_H-2))(v_H-1)+9(\mu-1)^2 v_H)\\+2-3v_I+3\mu(3+(\mu-3)\mu+2\theta(-1+\mu+v_H-\mu v_H))v_I\\-3(\mu-1)^2(3\mu v_H-2)v_I^2+5(\mu-1)^3 v_I^3\end{pmatrix}\dfrac{\kappa N^2}{s}$ |
| d) | $\kappa HN + \dfrac{1}{2}\begin{pmatrix}(\theta-2)\theta+(\mu-1)^2 v_I^2\\+\mu v_H(2+\mu(v_H-2))\end{pmatrix}\dfrac{\kappa N^2}{s}$ | $\kappa HN + \left(-\dfrac{1}{6\varepsilon}\right)\begin{pmatrix}17+2\varepsilon^3+\theta^3-18\mu+3\mu^3+3\theta^2(\mu-1)(v_I-1)+3\theta(-2+(\mu+v_I-\mu v_I)^2)\\+3(\mu-1)(12+(\mu-10)\mu)v_I-15(\mu-2)(\mu-1)^2 v_I^2\\+9(-1+\mu)^3 v_I^3+3\varepsilon^2(5+\theta-3\mu+3(-1+\mu)v_I)\\+3\varepsilon(9+2\theta-8\mu-2\theta\mu+\mu^2+2(7+\theta-4\mu)(\mu-1)v_I+7(\mu-1)^2 v_I^2)\end{pmatrix}\dfrac{\kappa N^2}{s}$ |

where $\varepsilon = -1 + \mu v_H + (1-\mu)v_I$. One can verify that when $0 \leq v_H, v_I \leq 1$, there is $-1 \leq \varepsilon \leq 0$.

Up till now, we have already established work schedule choice model and the net benefit $\delta_J^i$ can be derived readily with the equilibrium commuting cost and individual output. The higher the net benefit of a work schedule is, the more commuters switch to that schedule.

To better investigate travelers' decision-making process between the two work schedules, we now define $D_J(v_H, v_I)$ with respect to $v_H, v_I \in [0,1]$, to represent the net benefit difference in the group $J(=H, I)$ between normal schedule and flextime, where

$$D_J(v_H, v_I) = \delta_J^r - \delta_J^f. \quad (20)$$

Evidently, when $D_J(v_H, v_I) > 0$, adopting normal schedule can enjoy higher net benefit and commuters in group $J$ will stick to normal schedule $(v_J \nearrow)$. On the contrary, when $D_J(v_H, v_I) < 0$, it is more beneficial to adopt flextime and commuters in group $J$ will switch to flextime $(v_J \searrow)$. From Table 1, multiple cases and therewith multiple forms of $D_J(v_H, v_I)$ exist and $D_J(v_H, v_I)$ is indeed a piecewise function with respect to $v_H$ or $v_I$. For instance, for $\theta + \mu < 1$, only Case a) and b) occur. Thus, $D_J(v_H, v_I) = \delta_J^r - \delta_J^f = (F^r - P_J^r) - (F^f - P_J^f)$ first follows Case b) when $0 \leq v_I < \frac{\theta}{1-\mu}$ and later follows Case a) when $\frac{\theta}{1-\mu} \leq v_I \leq 1$. Similarly, $D_J(v_H, v_I)$ for $\theta + \mu > 1$ can also be derived.

To solve the equilibrium number of commuters adopting flextime, it suffices to find the stable points within the virtual $1 \times 1$ unit square, where $v_H$ and $v_I$ are the two axis. However, the stability condition and solutions vary depending on the locations of the equilibrium points in the virtual square (interior, corner or boundary), which will be examined respectively in what follows.

*4.2 Interior solutions*

For any interior solutions, i.e., both work schedules are used in both groups of commuters at equilibrium, the following lemma and proposition holds.

***Lemma 4.1*** *Assume $v_A, v_B \in (0,1)$. Point $(v_H, v_I) = (v_A, v_B)$ reaches stable equilibrium, if and only if, it satisfies that, $D_J(v_A, v_B) = 0$ and $\frac{\partial D_J(v_A, v_B)}{\partial v_J} < 0$, where $J = H, I$.*

Lemma 4.1 can be proved by contradiction. The detailed proof is omited here to save space but is available from the authors on request. It shows the stability conditions for interior equilibrium.

***Proposition 4.1*** *When $\theta \in (0, 1-\mu]$, there is no interior equilibrium solution, i.e., it is impossible to have $N_H^r, N_H^f, N_I^r, N_I^f > 0$.*

**Proof.** See Appendix D.

Later we will further show in Corollary 4.2 that when $\theta \in (0, 1-\mu]$, the equilibrium can only be obtained at the two corners of the unit square. At least one group of commuters all adopt the same type of work schedule.

*4.3 Corner solutions*

For the corner solutions, i.e., each groups of commuters use a single same work schedule at equilibrium, the following lemma holds.

***Lemma 4.2*** *Assume binary variables $v_A, v_B \in \{0,1\}$. Point $(v_H, v_I) = (v_A, v_B)$ reaches stable equilibrium, if and only if, it satisfies that*

$$\begin{cases} (v_J - 0.5) D_J(v_A, v_B) > 0 \ (J = H, I); \\ \text{or } D_J(v_A, v_B) = 0 \text{ and } \dfrac{\partial D_J(v_A, v_B)}{\partial v_J} < 0 \ (J = H, I), \ \forall v_A, v_B \in \{0,1\}. \end{cases} \quad (21)$$

**Proof.** The first row of the condition above is equivalent to $D_J(v_A, v_B) > 0$ if $v_J = 1$; and $D_J(v_A, v_B) < 0$ if $v_J = 0$ $(J = H, I)$, which guarantee the stability following Eq.(7). The second row covers the occasions when $\delta_J^i$ becomes equal coincidently at the corner, and its stability conditions should be similar to Lemma 4.1. □

Obviously, there are four possible corner solutions, where $(v_H, v_I) = (0,0), (0,1), (1,0), (1,1)$. The following proposition summarizes occurrence conditions of corner solutions.

***Proposition 4.2*** *The equilibrium is obtained at the corner:*

(1). At $(v_H, v_I) = (1,1)$: if and only if $\rho < 0.5$ $\left(\text{i.e., } \beta < \dfrac{1}{2} \kappa N\right)$;

(2). At $(v_H, v_I) = (0,0)$ : if and only if $\theta \in [0, 2(1-\mu))$ or $\theta \in [2(1-\mu), 1]$ and

$$\rho < \rho^{d*}, \text{ where } \rho^{d*} = \frac{\theta^3 - 3\theta^2\mu + 3\theta(\mu^2 + 2\mu - 1) + 3(1-\mu)^2(1+\mu)}{12(\theta + 2\mu - 2)}.$$

(3). Never at $(v_H, v_I) = (0,1), (1,0)$.

**Proof of Proposition 4.2.** See Appendix E.

The occurrence domain of stable equilibrium $(v_H, v_I) = (0,0)$ with respect to $\mu$ and $\theta$ is also shown in Figure 3, where Corollary 4.1 is obtained.

**Corollary 4.1** *When $\rho \leq \frac{1}{3}$, the situation in which all commuters use flextime can always be realized at equilibrium.*

In Proposition 4.2, the stability of the equilibrium at four corner cases is examined. Foremost, Proposition 4.2 (1) implies that when all firms adopt normal schedule initially, in the cases with relatively small unit time scheduling penalty, firms no longer have incentives to switching their offered work schedule to flextime, regardless of the proportions of household employees and the length of the *flex-interval*. In Proposition 4.2 (2) and Corollary 4.1, it is also found that the equilibrium can be attained in most cases when all firms adopt flextime unless there are long *flex-interval*, high household proportion and high scheduling penalty. Lastly, Proposition 4.2 (3) indicates that it is impossible to reach equilibrium at the other two corners without other policy intervention. The agglomeration of work schedule of any group of commuters (household or individual) has attracted the commuters in the other group to select that work schedule as well so as to maximize their individual net benefit.

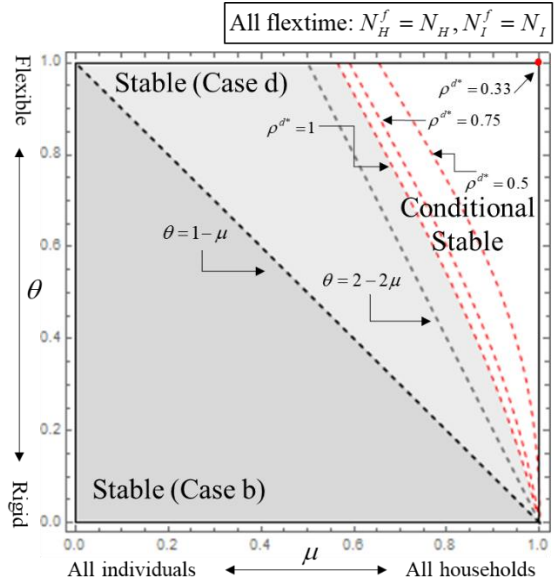

*Figure 3 The occurrence domain of stable equilibrium with respect to $\mu$ and $\theta$.*

### 4.4 Boundary solutions

Lastly for the boundary solutions, i.e., one group of commuters all uses the same work schedule whereas the other group does not, we have the following lemma.

***Lemma 4.3*** *Assume $v_A$ is a binary variable $\{0,1\}$ and $v_B \in [0,1]$. Point $(v_H, v_I) = (v_A, v_B)$ (or $(v_B, v_A)$) reaches stable equilibrium, if and only if, it satisfies the following two conditions.*

1.
$$\begin{cases} (v_A - 0.5) D_H(v_A, v_B) > 0 \ \left( \text{or } (v_A - 0.5) D_I(v_B, v_A) > 0 \right); \\ \text{or } D_H(v_A, v_B) = 0 \text{ and } \frac{\partial D_H(v_A, v_B)}{\partial v_H} < 0 \ \left( \text{or } D_I(v_B, v_A) = 0 \text{ and } \frac{\partial D_I(v_B, v_A)}{\partial v_I} < 0 \right), \forall v_A \in \{0,1\}; \end{cases}$$

2.
$$D_I(v_A, v_B) = 0 \text{ and } \frac{\partial D_I(v_A, v_B)}{\partial v_I} < 0 \ \left( \text{or } D_H(v_B, v_A) = 0 \text{ and } \frac{\partial D_H(v_B, v_A)}{\partial v_H} < 0 \right), \forall v_A \in \{0,1\}.$$

**Proof.** Condition one is in the similar form of Eq.(21) to assure the equilibrium points fall on the boundaries and condition two is in the same form of Lemma 4.1 to determine the exact location of the equilibrium point. Note that when $\nu_B = 0, 1$ at extreme cases, it is equivalent to the corner equilibrium. □

Here we focus on the possible solutions on the four boundaries. The occurrence conditions of boundary solutions are summarized in the following proposition.

**Proposition 4.3** *When $\theta \in (0, 1-\mu]$, the equilibrium can never be obtained on the boundaries, unless at the corners.*

**Proof.** See Appendix F. □

Proposition 4.3 shows that under certain conditions such that the length of *flex-interval* is too short to cover the arrival period of all individual employees, the equilibrium cannot be obtained when one group of commuters all uses the same work schedule and the other group does not. In fact, combining Proposition 4.1-4.3, the following corollary holds.

**Corollary 4.2** *When $\theta \in (0, 1-\mu]$, equilibrium can only be obtained when all travelers adopt the same work schedule (either normal schedule or flextime).*

Corollary 4.2 indicates that when flextime is not flexible enough, there is no distinct advantage of one schedule over the other. Therefore, various groups of commuters show no preference in selecting one specific schedule and they only have the inclinations towards the same work schedule in order to maximize their individual net benefit.

So far, the long-run equilibrium solutions are analytically obtained under some conditions. The more general equilibrium is also illustrated later in Section 6 with numerical examples.

5. **System optimum**

In Section 3 and 4, we have examined both the short-run and long-run equilibrium. Hereto from the perspective of city regulators, we turn to evaluate the economic efficiency of different flextime proportions to minimize total commuting cost and maximize total net benefit.

*5.1 Total commuting cost minimization*

Regarding short-run equilibrium, the total commuting cost is calculated as the sum of individual commuting cost, as

$$TC = \sum_{J}\sum_{i} N_J^i P_J^i, \text{ for } J = H, I \text{ and } i = r, f, \tag{22}$$

which is to be minimized so as to achieve short-run optimum.

Given $\mu$,

Table 4 summarizes the total commuting cost $TC$, the first-order derivative of $TC$ with respect to $v_H, v_I, \theta$ and the minimum $TC$ in all the four patterns depicted in Figure 2. The following lemma and proposition are derived from Table 4.

**Lemma 5.1** There is $\frac{\partial TC}{\partial v_I} \geq 0$ and $\frac{\partial TC}{\partial \theta} \leq 0$ for all cases and $\frac{\partial TC}{\partial v_H} \geq 0$ in Case a),b),c). As for Case d), there is $\frac{\partial TC}{\partial v_H} = 0$ at $v_H = \frac{-2+\theta+2\mu}{4\mu}$ and $\frac{\partial^2 TC}{\partial v_H^2} > 0$.

**Proposition 5.1** The global minimum of $TC$ is obtained when $\theta = 1, v_I = 0$ and

$$v_H = \begin{cases} \dfrac{-2+\theta+2\mu}{4\mu} & \text{when } \mu \in \left(\dfrac{1}{2}, 1\right]; \\ 0 & \text{otherwise.} \end{cases}$$

Lemma 5.1 indicates that at total commuting cost minimization, $v_I$ reaches its minimum and $\theta$ reaches its maximum. As for $v_H$, it reaches its minimum in Case a),b),c). In other words, increasing the length of *flex-interval* $\Delta$ or encouraging more individuals to use flextime can always lower $TC$ and thereby reduce the negative externalities in congestion as expected. Encouraging more household employees to use flextime can also lower $TC$ in most cases. However, in Case d) when $\mu, \theta$ and $N_H^f$ are relatively large, increasing $N_H^f$ (reducing $v_H$) can lead to initial decrease and later increase of $TC$ with possible minimum at $v_H = \frac{-2+\theta+2\mu}{4\mu}$. Indeed, when there are already too many household flextime employees, further encouraging more households to use flextime leads to larger negative congestion externalities. This can be ascribed to the rigidity of school start time. While in this paper we focus on the work start time distribution and its effect in congestion alleviation, the congestion due to school trips cannot be reduced without the aid of other measures, e.g., road tolls. Such inevitable school trips' congestion prevents the total commuting cost from decreasing continuously for large $N_H^f$. Nevertheless, as shown in Proposition 5.1, the minimum of $TC$ in Case d) is the smallest after comparing the minimum $TC$ among all four cases.

*Table 4  Total commuting cost and the related derivatives in each pattern depicted in Figure 2.*

| Case | a) | b) | c) | d) |
|---|---|---|---|---|
| $TC$ | $\left(\begin{array}{c}1-\mu+2\mu^2\\+\varepsilon\theta\end{array}\right)\frac{2\beta N^2}{s}$ | $\left(\begin{array}{c}1-\mu+2\mu^2\\-\theta(1-\mu v_H)\\+(1-\mu)^2 v_I^2\end{array}\right)\frac{2\beta N^2}{s}$ | $\left(\begin{array}{c}\mu+\mu^2\\-\theta\mu(1-v_H)\\+(1-\mu)^2 v_I^2\end{array}\right)\frac{2\beta N^2}{s}$ | $\left(\begin{array}{c}2\mu^2 v_H^2\\+(2\mu-\theta\mu-2\mu^2)v_H\\+\mu-\theta\mu+\mu^2\\+(1-\mu)^2 v_I^2\end{array}\right)\frac{2\beta N^2}{s}$ |
| $\frac{\partial TC}{\partial v_H}$ | | $\mu\theta\frac{2\beta N^2}{s}$ | | $\mu(2-\theta-2\mu+4\mu v_H)\frac{2\beta N^2}{s}$ |
| $\frac{\partial TC}{\partial v_I}$ | $2(1-\mu)^2 v_I\frac{2\beta N^2}{s}$ | $\theta(1-\mu)\frac{2\beta N^2}{s}$ | $2(1-\mu)^2 v_I\frac{2\beta N^2}{s}$ | |
| $\frac{\partial TC}{\partial \theta}$ | $-(1-\mu v_H)\frac{2\beta N^2}{s}$ | $\varepsilon\frac{2\beta N^2}{s}$ | $-(1-v_H)\mu\frac{2\beta N^2}{s}$ | $-\mu(1+v_H)\frac{2\beta N^2}{s}$ |
| Min $TC$ | $(1-2\mu+3\mu^2)\frac{2\beta N^2}{s}$ | $2\mu^2\frac{2\beta N^2}{s}$ | $\mu(1+\mu)\frac{2\beta N^2}{s}$ | $\begin{cases}\frac{(-1+4\mu+4\mu^2)}{8}\frac{2\beta N^2}{s}\\ \mu^2\frac{2\beta N^2}{s}\end{cases}$ |

Note: $\varepsilon = -1+\mu v_H + (1-\mu)v_I$ and the Min $TC$ in Case d) depends on the values of $\mu$:

$$\begin{cases}\frac{(-1+4\mu+4\mu^2)}{8}\frac{2\beta N^2}{s} & \text{when } \mu>\frac{1}{2},\\ \frac{\mu^2 2\beta N^2}{s} & \text{otherwise.}\end{cases}$$

*5.2 Total net benefit maximization*

In terms of long-run equilibrium, we define the total net benefit as the sum of the net benefit of all travelers, as

$$TB = \sum_{J}\sum_{i} N_J^i \delta_J^i, \text{ for } J = H, I \text{ and } i = r, f, \tag{23}$$

which is summarized in Table 5 and is to be maximized at long-run optimum.

One may notice that the total net benefit shown in Table 5 can be all transformed to $TB = H\kappa N^2 + \frac{\kappa N^3}{s} g(v_H, v_I, \theta, \mu, \rho)$, where $g(\cdot)$ is a function of $v_H, v_I, \theta, \mu$ and $\rho$.

Obviously, as $\kappa, N, H$ and $s$ are usually given, maximizing $TB$ amounts to maximizing $g(\cdot)$.

*Table 5  Total net benefit in each pattern depicted in Figure 2.*

| Case | Total Net Benefit: $TB = H\kappa N^2 + \frac{\kappa N^3}{s} g(v_H, v_I, \theta, \mu, \rho)$ |
|---|---|
| a) | $H\kappa N^2 + \frac{\kappa N^3}{s}\left(\theta(\mu(v_H - v_I) + v_I - 1)(2\mu(v_H - v_I) + 2v_I - \rho) - (1 - \mu + 2\mu^2)\rho\right)$ |
| b) | $H\kappa N^2 + \frac{\kappa N^3}{s}\left(\frac{2}{3}\theta^3 + \theta^2(2\mu v_H - 1) + \frac{4}{3}(1-\mu)^3 v_I^3 + (1-\mu)^2 v_I^2(2\mu v_H - \rho - 1) - \theta(1 - \mu v_H)(2\mu v_H - \rho) - (1 - \mu + 2\mu^2)\rho\right)$ |
| c) | $H\kappa N^2 + \frac{\kappa N^3}{s}\begin{pmatrix}\frac{-1}{3} + \frac{4}{3}v_I^3 + 2\mu\left((-1+\theta)(-1+v_H) + (1+v_H)v_I^2 - 2v_I^3\right) + \frac{1}{3}\mu^3\left(4 - 4v_I^3 + 6v_H(-1 + v_I^2)\right) \\ + \mu^2\left(-3 + 2\theta(-1 + v_H)^2 + 4v_H(1 - v_I^2) + v_I^2(4v_I - \rho - 1) - \rho\right) + \mu(2v_I^2 - 1 + \theta - \theta v_I)\rho - v_I^2(1+\rho)\end{pmatrix}$ |
| d) | $H\kappa N^2 + \frac{\kappa N^3}{s}\begin{pmatrix}\frac{1}{2} + \frac{1}{6}\theta^3 + \frac{1}{2}\theta^2(1 + \mu(v_H - 1)) + \frac{1}{6}\mu^3\left(3 + v_H(3 + 5(-3 + v_H)v_H) + 12v_H v_I^2 - 8v_I^3\right) \\ + \frac{1}{2}\theta\left(-3 + \mu^2(-1+v_H)^2 + 2\mu(1 + v_H(-1+\rho) + \rho)\right) \\ + \frac{1}{3}v_I^2(4v_I - 3(1+\rho)) + \frac{1}{2}\mu\left(v_H(1 + 4v_I^2 - 4\rho) - 1 - 2\rho + 4v_I^2(1 - 2v_I + \rho)\right) \\ - \frac{1}{2}\mu^2\left(1 + v_H(2 + 8v_I^2 - 4\rho) + 2\rho + 2v_I^2(1 - 4v_I + \rho) + v_H^2(-5 + 4\rho)\right)\end{pmatrix}$ |

**Lemma 5.2** *When $\theta \in (0, 1-\mu]$, the total net benefit decreases with $\rho$ (or $\beta$), ceteris paribus.*

**Proof.** See Appendix G.

Apparently, larger unit time scheduling penalty leads to higher cost in travelling to both workplace and school, and further results in the lower total net benefit. Only in the extreme case when all travelers are flextime individuals with long enough *flex-interval* (i.e., $\mu = 0, \theta = 1, v_I = 0$), the total net benefit is irrelevant to $\rho$.

Besides, with given $\theta, \mu, \rho$, $g(\cdot)$ can be reduced to the function of $(v_H, v_I)$. From the second partial derivative test, any local maximum of $g(\cdot)$ within the domains at $(v_A, v_B)$ should satisfy the following three conditions.

$$\begin{cases} \left.\dfrac{\partial g}{\partial v_H}\right|_{(v_H,v_I)=(v_A,v_B)} = 0, \left.\dfrac{\partial g}{\partial v_I}\right|_{(v_H,v_I)=(v_A,v_B)} = 0; \\ \left.\dfrac{\partial g}{\partial v_H^2}\right|_{(v_H,v_I)=(v_A,v_B)} < 0; \\ \det\left(\mathrm{H}(v_A, v_B)\right) > 0, \end{cases} \quad (24)$$

where the first row determines the location of the critical points $(v_A, v_B)$ and the next two rows guarantee the negative definiteness of the Hessian $\mathrm{H}(\cdot, \cdot)$ at $(v_A, v_B)$.

By verifying the aforementioned optimum conditions within the domain of $(v_H, v_I)$ in the virtual square, there is Proposition 5.2 on the properties of long-run optimum.

**Proposition 5.2** *With given $\theta, \mu, \rho$:*

1. *when $\theta \in (0, 1-\mu]$, the long-run optimum is attained at*

$$\left(v_H^*, v_I^*\right) = \begin{cases} (1,1), & \rho \in \left(0, \dfrac{1}{2}\theta - \dfrac{1}{3}\theta^2\right); \\ (0,0), & \rho \in \left(\dfrac{1}{2}\theta - \dfrac{1}{3}\theta^2, 1\right); \end{cases} \quad (25)$$

2. *when $\theta \in (1-\mu, 1]$, if it further satisfies that $\mu \in \left(2(\sqrt{2}-1), 1\right]$, $\theta \in \left(2\sqrt{2(1-\mu)} + \mu - 1, 1\right]$ and $\rho \in \left[\dfrac{2\theta(1+\mu) - \theta^2 - (1-\mu)^2}{4(\theta + 2\mu - 2)}, 1\right)$, the long-run optimum can only be attained when $v_I^* = 0$ and*

$$v_H^* = 1 - \frac{5+\theta-8\rho+2\sqrt{5(1-\mu)^2-\theta^2+\theta(5-9\rho)-10(1-\mu)\rho+16\rho^2}}{5\mu}; \quad (26)$$

*and else, the long-run optimum is attained at* $(v_H^*, v_I^*) = (0,0)$ or $(1,1)$.

**Proof of Proposition 5.2.** See Appendix H.

Following Proposition 5.2, it should be noted for $\theta \in (0, 1-\mu]$ that when $\rho = \frac{1}{2}\theta - \frac{1}{3}\theta^2$, the total net benefits at $(0,0)$ and $(1,1)$ are equal. In addition, though there is optimum for $\theta \in (1-\mu, 1]$ at $v_H^* > 0$ as shown in Proposition 5.2, it can be verified that $v_H^*$ in Eq.(26) reaches its maximum 0.07 at $\rho = 1$, $\mu = 1$ and $\theta = 1$. It suffices to say that long-run optimum in this case is obtained when flextime becomes the mainstream. In fact, when comparing $g(\cdot)$ at the four corners, we also notice Corollary 5.1 as follows.

***Corollary 5.1*** *The* long-run *optimum cannot be obtained* at $(v_H, v_I) = (1,0), (0,1)$.

**Proof.** See Appendix I.

To maximize $TB$, Proposition 5.2 suggest the optimal flextime proportions of household and individual travelers. For flextime with low flexibility, it is implied in Proposition 5.2 (1) that all travelers should agglomerate in their work schedule choice to reach long-run optimum. Whether to agglomerate towards normal schedule or flextime depends on the scheduling penalties $\rho$. For small $\rho$, the gains in reduced negative congestion externalities are smaller than the losses in reduced positive production externalities when more commuters adopt flextime. Namely, normal schedule brings more benefits than flextime at that $\rho$. Likewise, for large $\rho$, it shows a strong flextime tendency at long-run optimum due to the superiority of flextime. As flexibility in flextime increases, the optimum is still attained usually with work schedule agglomeration for all commuters. Only if there is large household proportion, long *flex-interval*, and high unit time scheduling penalty (i.e., the conditions listed in Proposition 5.2 (2)), it reaches the optimum at the majority of households (and all individuals) in flextime. In addition, Corollary 5.1 also shows that the long-run optimum cannot be reached when the household and individual commuters agglomerate and polarize at different work schedules, respectively. To maximize total net benefit, the agglomeration effects on productivity make it beneficial for most employees using a same schedule, regardless of their trip purposes.

One may find that the long-run optimum does not always coincide with the long-run equilibrium. For instance, when $\theta \in (0, 1-\mu]$, while the equilibrium can be attained at $(v_H, v_I) = (1,1) \& (0,0)$, the long-run optimum can be attained at either $(0,0)$ or $(1,1)$ depending on the value of $\rho$. To shed some light on the difference between equilibrium and system optimum, we next take a closer look at the scenarios when $\theta \in (0, 1-\mu]$ and $\rho \in (0, 0.5)$. A benchmark case at $(v_H, v_I) = (1,1)$ where all commuters use normal schedule is further considered to compare with long-run optimum. As discussed in Proposition 4.2, for $\rho < 0.5$, the benchmark case has achieved stable equilibrium already.

From Proposition 5.2, when $\theta \in (0, 1-\mu]$ and $\rho \in (0, 0.5)$, the percentage change of $TB$, $\eta$, is defined as

$$\eta = \frac{TB|_{SO} - TB|_{v_H=1, v_I=1}}{\left| TB|_{v_H=1, v_I=1} - H\kappa N^2 \right|} = \begin{cases} 0, & \text{if } \rho \in \left(0, \frac{1}{2}\theta - \frac{1}{3}\theta^2\right); \\ \frac{\left((2\theta-3)\theta + 6\rho\right)\theta}{6(2\mu^2 - \mu + 1)\rho}, & \text{if } \rho \in \left(\frac{1}{2}\theta - \frac{1}{3}\theta^2, \frac{1}{2}\right); \end{cases} \quad (27)$$

where $TB|_{v_H=1, v_I=1} = H\kappa N^2 + \frac{\kappa N^3}{s}(-1 + \mu - 2\mu^2)\rho$ reaching maximum at $\mu = 0.25$ and $H\kappa N^2$ is the identical constant term of productivity in all $TB$ as in Table 5. It can be readily verified that $\text{Min}(\eta) = 0$ when $\rho < \frac{1}{2}\theta - \frac{1}{3}\theta^2$ or $\theta = 0$ and $\text{Max}(\eta) = \frac{2}{3}$ when $\rho = 0.5$, $\theta = 1$ and $\mu = 0$. In other words, by driving the system from user equilibrium to long-run optimum, $TB$ can increase up to 66.7% when leaving out the constant term of productivity.

### 5.2.1 Pigouvian policy

Given the possible increase of $TB$ from equilibrium to optimum, it is straightforward to explore the managing measures to reach the long-run optimum naturally as stable equilibrium from the benchmark case. Here, the simple Pigouvian tax/subsidy policy will be examined. Suppose that in Pigouvian policy, an amount of tax/subsidy $\sigma_H, \sigma_I$ is imposed/provided to flextime household and individual commuters differentially, where positive (negative) $\sigma$ indicates Pigouvian tax (subsidy) for flextime employees. The updated net benefit difference between normal schedule and flextime for each group of commuters become $D_J^p(v_H, v_I) = D_J(v_H, v_I) + \sigma_J$ $(J = H, I)$. The long-run optimum at equilibrium implies that there exists $\sigma_H^*, \sigma_I^*$ allowing $D_J^p(v_H^*, v_I^*)$ to satisfy Lemma 4.1-4.3 to achieve stable equilibrium from the benchmark case $(v_H, v_I) = (1,1)$. In Section 6, the feasibility of Pigouvian policy is illustrated numerically. Nonetheless, it should be noted that Pigouvian policy may not always be effective and further examination is needed on its feasibility conditions.

### 6. **Numerical example**

In this section, the long-run equilibrium and optimum as well as the Pigouvian policy are exemplified.

We first illustrate the long-run equilibrium conditions numerically. Let $s = 100$ veh/min, $H = 400$ min, $N = 10000$ and $\kappa = 0.001$. Different values of $\beta, \theta$ and $\mu$ can lead to different equilibrium. In

Figure 4, the 3D plots of the net benefit difference $D_J(v_H, v_I)$ $(J = H, I)$ and their contour plots at $D_J(v_H, v_I) = 0$ are shown with given $\rho, \theta$ and $\mu$. Note that in the 3D plot, the blue and orange surfaces represent household and individual travelers, respectively. The plane in red is the reference plane at 0. Thus, as discussed earlier, when $D_J(v_H, v_I) > 0$, normal schedule is more attractive than flextime to group $J$ commuters, and vice versa. Furthermore, the solid lines in blue and orange in the contour plot represent the contours of household and individual travelers at $D_J(v_H, v_I) = 0$, respectively. The thick dashed lines in blue and orange represent the unilateral equilibrium points of household and individual travelers. Their intersections, which are shown in red dots, are the actual equilibrium points.

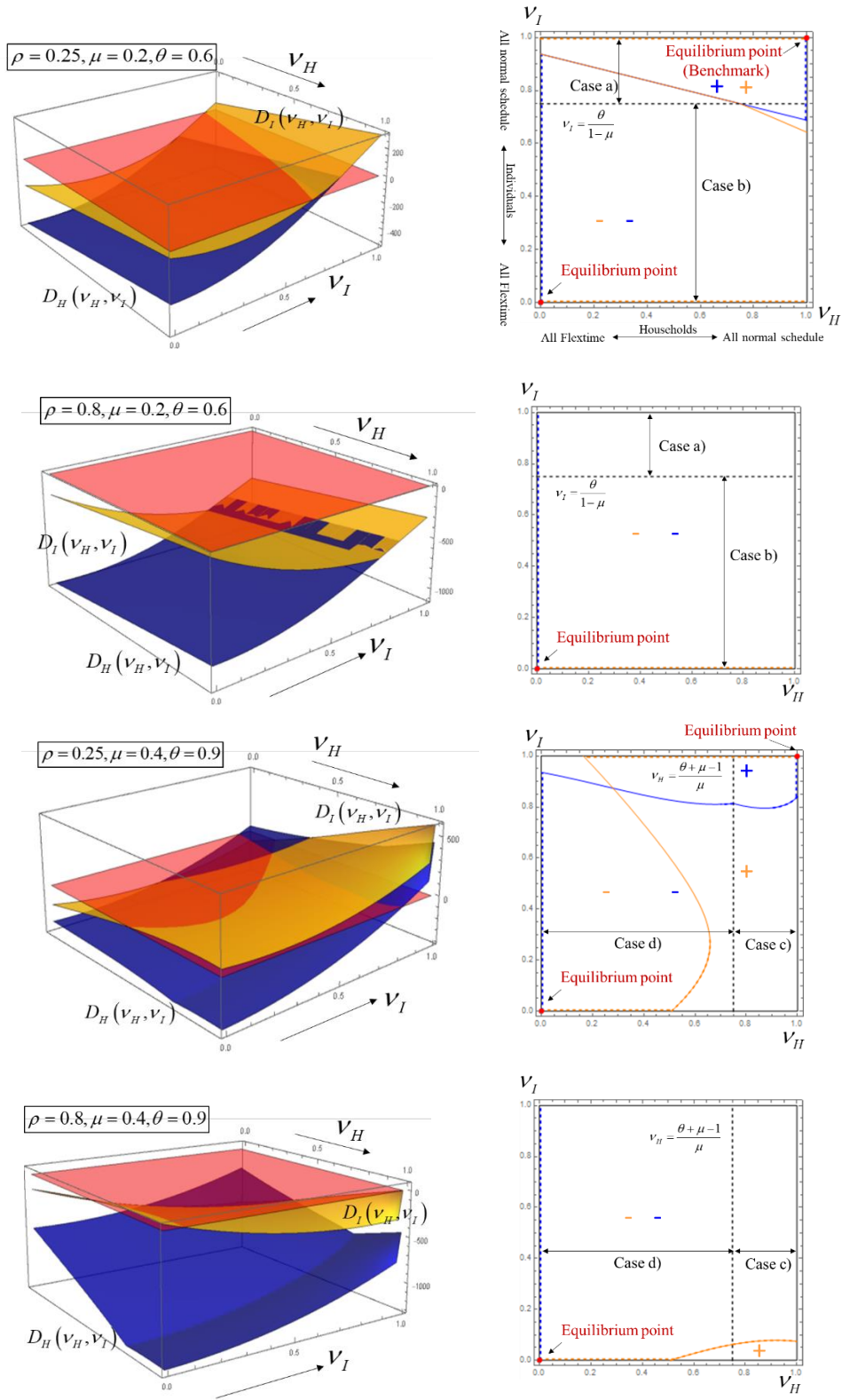

*Figure 4 The 3D plots and contour plots at $D_J(v_H, v_I) = 0$ of the net benefit difference between normal schedule and flextime.*

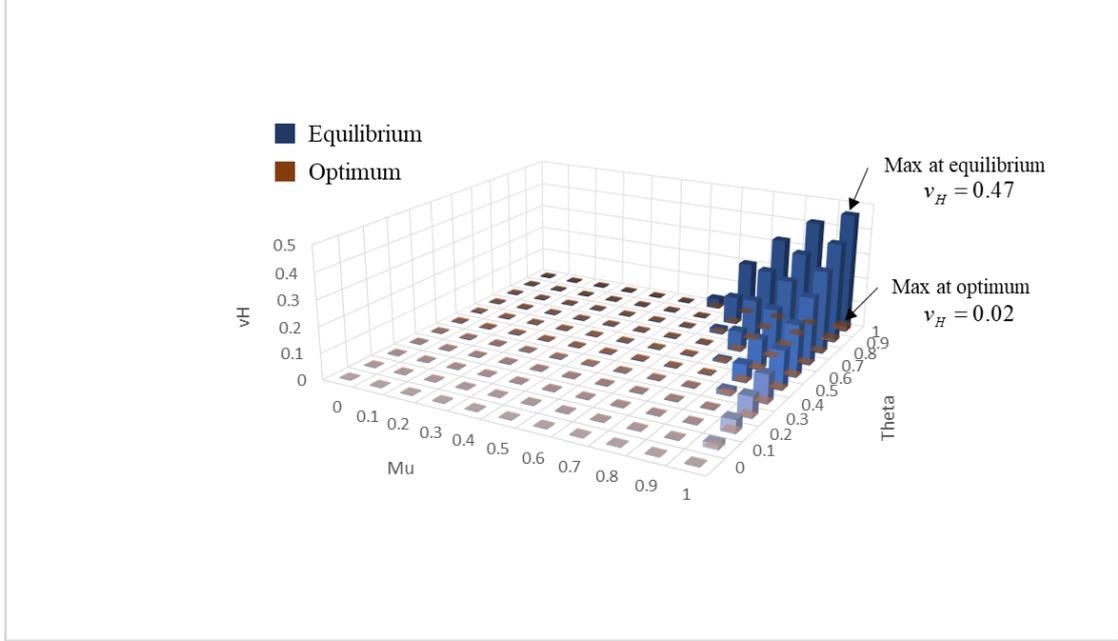

*Figure 5 The equilibrium and optimum $v_H$ with respect to $\mu, \theta$ at $\rho = 0.8$, where $v_I = 0$ at both equilibrium and optimum.*

In Figure 4, for $\mu = 0.2$ and $\theta = 0.6$, it is Case a) and b) with $\theta + \mu < 1$. We can find out that at $\rho = 0.25 < 0.5$ (i.e., $\beta = 2.5$), there are two possible equilibrium $(v_H, v_I) = (1,1), (0,0)$. At $\rho = 0.8 > 0.5$ ($\beta = 8$), $(1,1)$ is no longer stable and there is only one equilibrium $(0,0)$. As for $\mu = 0.4$ and $\theta = 0.9$, it is Case c) and d) with $\theta + \mu > 1$. Likewise, there are two stable equilibriums $(1,1), (0,0)$ at $\rho = 0.25$ and only one stable equilibrium $(0,0)$ at $\rho = 0.8$.

We further enumerate the user equilibrium points with different $\theta, \mu \in [0,1]$ in step size 0.1 at two different $\rho = 0.25 \,\&\, 0.8$. At $\rho = 0.25$, it is found that the possible equilibriums are $(v_H, v_I) = (1,1) \,\&\, (0,0)$ only for all $\theta, \mu \in [0,1]$. At $\rho = 0.8$, the equilibrium is $(v_H, v_I) = (0,0)$ in most cases as shown in Figure 5. Only with large $\mu, \theta$, the equilibrium is attained when all individuals use flextime $(v_I = 0)$, and some households choose normal schedule. it is worth noting that for the two $\rho = 0.25 \,\&\, 0.8$, there is interior equilibrium neither at $\theta + \mu < 1$ as proofed in Proposition 4.1, nor at $\theta + \mu \geq 1$ from observations. Based on the results, Proposition 4.1-4.3 are verified.

Next, we illustrate the long-run optimum in *TB* maximization. The value of $s, H, N, \kappa$ remains unchanged. At different $\theta, \mu \in [0,1]$ with step size 0.1, the optimal $v_H, v_I$ are enumerated. Compared to the equilibrium result, the optimum at $\rho = 0.25$ for all $\theta, \mu \in [0,1]$ locates at $(0,0)$ only. As for the optimum at $\rho = 0.8$ (presented in Figure 5), one can see that only when $\mu$ is large enough, the system optimum is reached with $v_H > 0$. For $\theta \in (1-\mu, 1]$, it can be derived from Proposition 5.2 that only when $\mu > 0.95$, it is possible to have optimum at $v_H^* > 0$ at $\rho = 0.8$, even though Max $v_H^* < 0.02$. Evidently, we can observe the optimum with $v_H > 0$ rarely at $(\mu, \theta) = (1,1)$ in this example. Proposition 5.2 is verified.

On top of equilibrium and optimum results, the Pigouvian policy is now explored. From Section 5.2.1, identifying $\sigma_H^*, \sigma_I^*$ suffices to translate the surfaces of net benefit difference (referred to Figure 4) upward/downward in the 3D plot to attain stable equilibrium at the long-run optimum $(v_H^*, v_I^*)$. Therefore, take the case when $\rho = 0.25$, $\mu = 0.2$ and $\theta = 0.6$ as an example (Figure 4), it can be derived that with $\sigma_H^*, \sigma_I^* \in (-\infty, -300)$, the optimum $(v_H, v_I) = (0,0)$ becomes the only equilibrium point with $\eta = 19.09\%$ compared to the benchmark case. In other words, subsidizing employees to use flextime can make the all-flextime optimum desirable at equilibrium. With differentiated Pigouvian policy, both groups of employees all switch to flextime voluntarily from the all-normal-schedule benchmark and long-run optimum is achieved in this case. In fact, it is noteworthy that while we only consider the tax/subsidy for flextime commuters, $\sigma_H^*, \sigma_I^*$ can be offset readily to achieve various tax and subsidy combinations for different groups' commuters in different work schedule.

## 7. Concluding remarks

To sum up, in this study a bi-level economic model is developed to examine the interrelationship between bottleneck congestion and commuters' work schedule choice behavior. In addition to encapsulating negative congestion externalities and positive production externalities, we extend the model to explicitly consider the heterogenous commuters with home-school-work and home-work trip purposes, i.e., the household and individual travels. The tradeoffs of commuters with different trip purposes between commuting cost and productivity are examined at length.

Unlike the conventional individual commuters, household commuters need to finish the school trip first followed by the work trip. In the short-run, when commuters cannot alter their work as well as the work schedule, the commuting costs of normal schedule commuters are generally higher than those in flextime at equilibrium for all trip purposes. In the long-run, some equilibrium properties are explored when employers and employees can reselect their work and work schedule. For instance, in the case with small $\rho(<0.5)$, commuters have no incentives to switch to firms in flextime from the initial normal schedule, regardless of the household proportions and the length of the *flex-interval*. In addition, with short *flex-interval* ($\theta \in (0, 1-\mu]$), all commuters agglomerate in work schedule choice at equilibrium.

We further scrutinize the system optimum at both short-run to minimize total commuting cost and at long-run to maximize the total net benefit. We notice that the rigid school schedule for household commuters can impede the benefits of flextime in total commuting cost saving in case of large household proportion. As such, the short-run optimum is attained with longest *flex-interval*, all flextime individuals and a specific proportion of flextime households. For long-run optimum, it is also found that all commuters should agglomerate in one work schedule mostly. Under certain circumstances, differentiated Pigouvian policy can turn the optimum into stable equilibrium from the initial benchmark case. Nonetheless, due to the agglomeration effect, the long-run optimum can never be reached when commuters' work schedule choices polarize, i.e., when all households choose one work schedule while all individuals choose the other.

Besides, some extensions are expected to be further addressed in future studies. Following Section 5.2.1, the differentiated Pigouvian policy against employees with multiple trip purposes should be explored in a more detailed form, so that its potential merit in maximizing total net benefit can be fully recognized. Future work could also include testing the assumptions in this model against user behavior via stated/revealed preference surveys. Moreover, in addition to temporal effects, it would be interesting to consider the spatial effects in work schedule choice with household travelers, particularly when telecommuting is widely available. Lastly, it is possible to extend the modeling framework to consider more general shared trip purposes (e.g., ridesharing in the family) in later work.

## Appendix A.  List of notations

*Parameters*

| | |
|---|---|
| $N$ | Total number of commuters. |
| $N_H, N_I$ | Equilibrium number of household and individual commuters. |
| $\mu$ | Proportion of household commuters, $\mu \in [0,1]$. |
| $t^s$ | Uniform school start time. |
| $t^w$ | Work start time of normal(rigid) schedule employees. |
| $\alpha, \beta, (\rho)$ | Value of time, (relative) unit time scheduling penalty of early arrival |
| $s$ | Constant service capacity of the bottleneck. |
| $H$ | Total hours of work per day. |
| $\Delta$ | Length of the *flex-interval* for flextime commuters. |
| $\theta$ | Length ratio of the actual *flex-interval* to the longest *flex-interval*, $\theta = \dfrac{\Delta}{N/s} \in (0,1]$. |

*Variables*

| | |
|---|---|
| $N^r, N^f$ | Equilibrium number of commuters in normal schedule and flextime. |
| $N^i(t)$ | The number of employees on duty in work schedule $i(=r,f)$ at time $t$. |
| $v_H, v_I$ | Proportions of normal schedule employees among household and individual commuters respectively, $v_H, v_I \in [0,1]$. |
| $N_J^i$ | Equilibrium number of $J(=H,I)$ commuters in work schedule $i$. |
| $\pi^i, F^i, \omega^i$ | The profit, output and wage per employee per day of firms in work schedule $i$. |
| $U_J^i$ | Individual net income of $J$ commuters in work schedule $i$. |
| $P_J^i$ | Aggregated travel cost of group $J$ commuters in work schedule $i$. |
| $\delta_J^i$ | Net benefit per employee of group $J$ commuters in work schedule $i$. |
| $\kappa$ | A constant parameter in the instantaneous output of employees. |
| $N^{fe}$ | Number of flextime commuters who arrive at workplace early before the start of *flex-interval* $t^w - \Delta$. |
| $T(t)$ | Travel time on the highway from home to workplace for commuters departing at $t$. |
| $r_J^i(t)$ | Flow rate of group $J$ commuters in work schedule $i$ entering the bottleneck at time $t$. |
| $p_J^i(t)$ | Generalized travel cost of group $J$ commuters in work schedule $i$ departing at $t$. |

**Appendix B.    Derivations of Table 2**

For Case a), the commuting cost in the morning

$$P_H^f\Big|_{\text{Morning}} = \beta\frac{N_H}{s} + \beta\left(\frac{N_H}{s} + t^w - \Delta - t^s\right) \quad, \quad P_H^r\Big|_{\text{Morning}} = \beta\frac{N_H}{s} + \beta\left(\frac{N_H}{s} + t^w - t^s\right) \quad,$$

$P_I^f\Big|_{\text{Morning}} = \beta(t^w - \Delta - t^s)$ and $P_I^r\Big|_{\text{Morning}} = \beta\frac{N_I}{s}$. To note, the first terms in $P_H^f\Big|_{\text{Morning}}, P_H^r\Big|_{\text{Morning}}$ are the commuting costs of the school trip and the second terms are the costs of work trip. For Case b), the household commuting costs $P_H^f\Big|_{\text{Morning}}, P_H^r\Big|_{\text{Morning}}$ are the same as those in Case a). As for individual commuters, $P_I^f\Big|_{\text{Morning}}$ is the same as

that in Case a), $P_I^r\big|_{\text{Morning}} = \beta\left(\dfrac{N_I^r}{s} + t^w - \Delta - t^s\right)$. For Case c), $P_H^f\big|_{\text{Morning}}, P_H^r\big|_{\text{Morning}}$ are the same as those in Case a) as well, and $P_I^f\big|_{\text{Morning}} = 0$, $P_I^r\big|_{\text{Morning}} = \beta\dfrac{N_I^r}{s}$. Lastly for Case d), $P_H^f\big|_{\text{Morning}}$ is the same as that in Case a), and

$$P_H^r\big|_{\text{Morning}} = \beta\left(\dfrac{N_H + N_H^r}{s} + t^w - \Delta - t^s\right) + \beta\left(t^w - t^s + \left(\dfrac{N_H + N_H^r}{s} + t^w - \Delta - t^s\right)\right),$$

while $P_I^f\big|_{\text{Morning}}, P_I^r\big|_{\text{Morning}}$ are the same as those in Case c). As the evening commuting cost is assumed to be the same as that in the morning, Table 2 is derived after mathematical manipulations.

**Appendix C.    Derivations of Table 3**

While the daily output of normal schedule commuters can be easily derived following Eq.(9), we focus on the output of flextime commuters. For flextime commuters, as shown in Eq.(10), the average individual output combines the output of both the arrivals earlier than the start of *flex-interval* $N^{fe}$, and the continuous arrivals within the *flex-interval* $N^f(t'') - N^f(t')$. Again, the commuters $N^{fe}$ start (and leave) work at the same time and the derivation of their output is similar to that in normal schedule. Attention is then paid to the continuous arrivals $N^f(t'') - N^f(t')$ and in what follows, the average output of arrivals during time interval $[t',t'']$ is derived.

As shown in Figure 6, there are continuous arrivals at workplace (work start) between $t'$ and $t''$ with arrival rate $s$, and hence their leaving from work is between $t'+H$ and $t''+H$ with identical leaving rate $s$. In a trial when the interval $[t',t'']$ is discretized into three unit sub-intervals, the individual output of arrivals at the first sub-interval is the sum output of zone A1, B1, C1, D and A2, where D is the identical output during $[t'',t'+H]$ shared by all arrivals within $[t',t'']$. Similarly, the output of the second interval is the sum output of zone B1, C1, D, A2 and B2 and lastly the output of the third interval is the sum output of zone C1, D, A2, B2 and C2. The total output is $(A1+3\times A2)+(2\times B1+2\times B2)+(3\times C1+C2)+3\times D$. In addition, it should be noted that the cumulative work-start commuters at any $t$ equal to the cumulative work-end commuters at $t+H$. Thus, after mathematical manipulations, we can derive the average output as $\dfrac{2(t''-t')(9N+2(t''-t')s)}{27}$. Likewise, when $[t',t'']$ is discretized into n sub-intervals, the average output of these commuters is $\dfrac{(1+n)(t''-t')(3nN+(n-1)(t''-t')s)}{6n^2}$. When $n\to+\infty$, there is

$$\frac{1}{N^f(t'')-N^f(t')}\int_{t'}^{t''}\left(N^{f\,\prime}(t^a)\int_{\omega}^{\omega+H}y(t)\,\mathrm{d}t\right)\mathrm{d}\omega = \frac{(t''-t')(3N+(t''-t')s)}{6}. \quad (28)$$

With Eq.(28), we can derive the average output of flextime commuters.

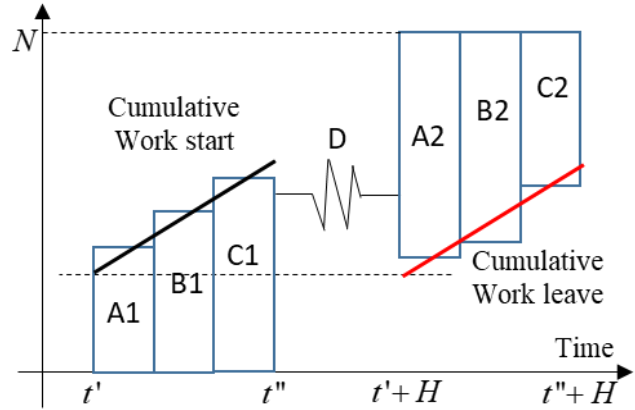

*Figure 6 The average output of continuous arrivals at workplace during time interval $[t',t'']$.*

**Appendix D.    Proof of Proposition 4.1**

When $\theta\in(0,1-\mu]$, only Case a) or b) may occur. For Case a), by solving

$D_J(v_H,v_I)=0$ ($J=H,I$), $\to v_I=\dfrac{1-2v_H\mu+\rho}{2-2\mu}$. Substituting the $(v_H,v_I)$ into

$\dfrac{\partial D_J(v_H, v_I)}{\partial v_J}$, there are $\dfrac{\partial D_H(v_H, v_I)}{\partial v_H} = \dfrac{2\theta\mu\kappa N^2}{s} > 0$ and

$\dfrac{\partial D_I(v_H, v_I)}{\partial v_I} = \dfrac{2\theta(1-\mu)\kappa N^2}{s} > 0$, which contradict with Lemma 4.1.

Likewise, for Case b), from $D_J(v_H, v_I) = 0$, there is $v_H = \dfrac{1-2\theta+\rho}{2\mu}, v_I = \dfrac{\theta}{1-\mu}$ and hence $\dfrac{\partial D_H(v_H, v_I)}{\partial v_H} = \dfrac{2\theta\mu\kappa N^2}{s} > 0$, which also contradicts with Lemma 4.1.

Thus, there is no internal equilibrium $v_H, v_I \in (0,1)$ when $\theta \in (0, 1-\mu]$.

**Appendix E.    Proof of Proposition 4.2**

For $(v_H, v_I) = (1,1)$, only Case a) and c) may occur. For Case a),

$D_H(1,1) = D_I(1,1) = \dfrac{\theta(0.5-\rho)\kappa N^2}{s}$. When $\rho < 0.5$, $D_H(1,1), D_I(1,1) > 0$. For Case

c), $D_H(1,1) = \dfrac{\theta(0.5-\rho)\kappa N^2}{s}$, $D_I(1,1) = \dfrac{(1-\mu)(0.5-\rho)\kappa N^2}{s}$. Again when

$\rho < 0.5$, $D_H(1,1), D_I(1,1) > 0$ and thus $(v_H, v_I) = (1,1)$ is stable.

For $(v_H, v_I) = (0,0)$, only Case b) and d) may occur. For Case b),

$D_H(0,0) = -\dfrac{(6-9\theta+4\theta^2+12\rho)\theta\kappa N^2}{6s}$, $D_I(0,0) = -\dfrac{(6-9\theta+4\theta^2)\theta\kappa N^2}{6s}$. One may

verify that $D_H(0,0), D_I(0,0) < 0$ for all $\theta \in (0, 1-\mu)$. As for Case d),

$D_H(0,0) = \dfrac{2(\theta+2\mu-2)(\rho-\rho^{d*})\kappa N^2}{s}$, where $\rho^{d*}$ is defined as in Proposition

4.2. When $\theta < 2-2\mu$, $\rho^{d*} < 0$ and hence $D_H(0,0) < 0$. When $\theta \geq 2-2\mu$, if

$\rho < \rho^{d*}$, $\to D_H(0,0) < 0$. In addition,

$D_I(0,0) = -\dfrac{\left(\theta^3 - 3\theta^2\mu + 3(1-\mu)^2(1+\mu) + 3\theta(\mu(2+\mu)-1)\right)\kappa N^2}{6s}$. It can be verified that $D_I(0,0) < 0$ for all $\theta \in (1-\mu, 1)$.

For $(v_H, v_I) = (1,0)$, Case b) and c) may occur. For Case b),

$D_H(1,0) = -\dfrac{\theta\left(4\theta^2 + 3\theta(4\mu-3) + 6(1-\mu)(1-2\mu+2\rho)\right)\kappa N^2}{6s(1-\mu)}$,

$D_I(1,0) = -\dfrac{\theta\left(6 - 9\theta + 4\theta^2 - 18\mu + 12\theta\mu + 12\mu^2\right)\kappa N^2}{6s(1-\mu)}$. It can be verified that the conditions when $D_H(1,0) > 0$ and $D_I(1,0) < 0$ cannot hold simultaneously. For

Case c), $D_H(1,0) = -\dfrac{(1-5\mu+4\mu^2+12\theta\rho)\kappa N^2}{6s}$, $D_I(1,0) = -\dfrac{(1-\mu)(1-4\mu)\kappa N^2}{6s}$.

Again $D_H(1,0) > 0$ and $D_I(1,0) < 0$ cannot hold simultaneously. Thus, equilibrium cannot be attained at $(v_H, v_I) = (1,0)$.

Lastly for $(v_H, v_I) = (0,1)$, Case a) and d) may occur. In Case a)

$D_H(0,1) = D_I(0,1) = \dfrac{\theta(1-2\mu-2\rho)\kappa N^2}{s}$. It is impossible to have $D_H(0,1) D_I(0,1) < 0$. For Case d)

$D_H(0,1) = -\dfrac{\begin{pmatrix}\theta^3 - 3\theta^2\mu + 3\theta(-1+\mu^2+\mu(2-4\rho)) \\ +(1-\mu)(2+5\mu^2+\mu(-7+24\rho))\end{pmatrix}\kappa N^2}{6\mu s}$, and

$D_I(0,1) = -\dfrac{\begin{pmatrix}\theta^3 - 3\theta^2\mu + 3\theta(-1+2\mu+\mu^2) \\ +(1-\mu)(2+5\mu^2+\mu(-7+12\rho))\end{pmatrix}\kappa N^2}{6\mu s}$. It can also be verified that $D_H(0,1) < 0$ and $D_I(0,1) > 0$ cannot hold simultaneously. Thus, equilibrium cannot be attained at $(v_H, v_I) = (0,1)$.

## Appendix F.   Proof of Proposition 4.3

For the boundary at $v_I = 0$, only Case b) may occur. From Lemma 4.3, if there is one equilibrium $(v_H, 0)$ on the boundary (except the corner point), it should satisfy that

$$D_H(v_H, 0) = 0, \frac{\partial D_H(v_H, 0)}{v_H} < 0 \text{ and } \begin{cases} D_I(v_H, 0) < 0 \\ D_I(v_H, 0) = 0, \frac{\partial D_I(v_H, 0)}{v_I} < 0 \end{cases}.$$ However, it

can be verified that there is no such $v_H \in (0,1)$. For the boundary at $v_H = 0$, both Case a) and Case b) may occur. If there is one equilibrium $(0, v_I)$ on the boundary (except the corner point), it should satisfy that $D_I(0, v_H) = 0$, $\frac{\partial D_I(0, v_I)}{v_I} < 0$ and

$$\begin{cases} D_H(0, v_I) < 0 \\ D_H(0, v_I) = 0, \frac{\partial D_H(0, v_I)}{v_H} < 0 \end{cases}.$$ Again, there is no such $v_I \in (0,1)$. For similar

reasoning, it also can be found that at the other boundaries at $v_I = 1$ (Case a)) and $v_H = 1$ (both Case a) and b)), no equilibrium is attained.

## Appendix G.   Proof of Lemma 5.2

When $\theta + \mu < 1$, only Case a) and b) may occur. In Case a), $\frac{\partial g}{\partial \rho} = -1 + \mu - 2\mu^2 - \varepsilon\theta$.

Again $\varepsilon = -1 + \mu v_H + (1-\mu)v_I$. Since $-1 \leq \varepsilon \leq 0$, $\frac{\partial g}{\partial \rho \partial \theta} \geq 0$, when $\theta = 1 - \mu$, we have $\text{Max} \frac{\partial g}{\partial \rho} = -(1+\varepsilon)(1-\mu) - 2\mu^2 \leq 0$, $\mu, \theta \in (0,1]$. Additionally, we have $\frac{\partial \rho}{\partial \beta} > 0$, from $\frac{\partial g}{\partial \rho} \leq 0$ we can get $\frac{\partial g}{\partial \rho} \leq 0$. In Case b),

$\frac{\partial g}{\partial \rho} = -(1+\varepsilon)(1-\mu) - 2\mu^2 + (1-\mu)^2(1-v_I)v_I$. Since $\frac{\partial g}{\partial \rho \partial v_I} = -2(1-\mu)^2 \leq 0$, when

$v_I = 0$, we have $\text{Max} \frac{\partial g}{\partial \rho} = -(1+\varepsilon)(1-\mu) - 2\mu^2 \leq 0$, $\mu, \theta \in (0,1]$ and thus $\frac{\partial g}{\partial \beta} \leq 0$.

**Appendix H.  Proof of Proposition 5.2**

Foremost, we determine if there is any local maximum for each case. When $\theta \in (0, 1-\mu]$, it is Case a) or b) and there is always $\frac{\partial g}{\partial v_H^2} = 4\theta\mu^2 \geq 0$ for all $v_H, v_I \in [0,1]$. Thus, there is no local maximum within the domains of those cases. Next, we compare $g(\cdot)$ at the boundaries. That is to compare $g(\cdot)|_{v_H=0, v_I=\frac{\theta}{1-\mu}}$, $g(\cdot)|_{v_H=1, v_I=\frac{\theta}{1-\mu}}$, $g(\cdot)|_{v_H=\frac{\mu+\theta-1}{\mu}, v_I=0}$, $g(\cdot)|_{v_H=\frac{\mu+\theta-1}{\mu}, v_I=1}$ and the other four corners in each case. It can be found that when $\theta \in (0, 1-\mu]$, the maximum $g(\cdot)$ can only be attained at the corners and Eq.(25) is derived.

When $\theta \in (1-\mu, 1]$, it can be verified that $\frac{\partial g}{\partial v_H^2} = 4\theta\mu^2 \geq 0$ in the domain of Case c). In other words, there is no local maximum within Case c). As for Case d), it can be found that there is one local maximum within the domain on the boundary of $v_I = 0$ for some values of $\theta, \mu, \rho$. Thus, we compare $g(\cdot)$ at the boundaries as well as the point at a local maximum. That is to compare $g(\cdot)|_{v_H=v_H^*, v_I=0}$, $g(\cdot)|_{v_H=0, v_I=\frac{\theta}{1-\mu}}$, $g(\cdot)|_{v_H=1, v_I=\frac{\theta}{1-\mu}}$, $g(\cdot)|_{v_H=\frac{\mu+\theta-1}{\mu}, v_I=0}$, $g(\cdot)|_{v_H=\frac{\mu+\theta-1}{\mu}, v_I=1}$ and the other four corners. When there exists a local maximum, it also reaches the maximum $g(\cdot)$ for all $v_H, v_I \in [0,1]$. This completes the proof.

## Appendix I. Proof of Corollary 5.1

If $(v_H, v_I) = (0,1)$ is the long-run optimum, apparently there are $g(\cdot)|_{v_H=1, v_I=1} < g(\cdot)|_{v_H=0, v_I=1}$ and $g(\cdot)|_{v_H=0, v_I=0} < g(\cdot)|_{v_H=0, v_I=1}$, $\exists \theta, \mu, \rho \in (0,1)$. When $\theta + \mu < 1$, with $g(\cdot)|_{v_H=1, v_I=1} - g(\cdot)|_{v_H=0, v_I=1} = 2\mu\theta(1-\mu-\rho)$ and

$$g(\cdot)|_{v_H=0, v_I=0} - g(\cdot)|_{v_H=0, v_I=1} = \frac{(\theta(2\theta-3) + 6(1-\mu)(\mu+\rho))\theta}{3}$$, there is no satisfied $\theta, \mu, \rho$. When $\theta + \mu \geq 1$, we cannot find such $\theta, \mu, \rho$ neither with

$$g(\cdot)|_{v_H=1, v_I=1} - g(\cdot)|_{v_H=0, v_I=1} = -\frac{1}{6}\begin{pmatrix} \theta^3 + 3\theta^2(1-\mu) + 5(1-\mu)^3 \\ + 3\theta(-3 + \mu(2 + \mu + 4\rho)) \end{pmatrix}$$

and

$$g(\cdot)|_{v_H=0, v_I=0} - g(\cdot)|_{v_H=0, v_I=1} = \frac{(1-\mu)^2(-1 + 4\mu + 6\rho)}{3}.$$

If $(v_H, v_I) = (1,0)$ is the system optimum, obviously there are $g(\cdot)|_{v_H=1, v_I=1} < g(\cdot)|_{v_H=1, v_I=0}$ and $g(\cdot)|_{v_H=0, v_I=0} < g(\cdot)|_{v_H=1, v_I=0}$, $\exists \theta, \mu, \rho \in (0,1)$. Similarly, there is no such $\theta, \mu, \rho$ to satisfy the conditions above. Thus, $(v_H, v_I) = (0,1) \text{ or } (1,0)$ cannot be the long-run optimum.

## Reference

Arnott, R. (2007). Congestion tolling with agglomeration externalities. *Journal of Urban Economics, 62*(2), 187-203. doi:http://dx.doi.org/10.1016/j.jue.2007.03.005

Arnott, R., de Palma, A., & Lindsey, R. (1990). Economics of a bottleneck. *Journal of Urban Economics, 27*(1), 111-130. doi:http://dx.doi.org/10.1016/0094-1190(90)90028-L

Arnott, R., Rave, T., & Schöb, R. (2005). Alleviating urban traffic congestion. *MIT Press Books, 1*.

Ben-Akiva, M., Cyna, M., & de Palma, A. (1984). Dynamic model of peak period congestion. *Transportation Research Part B: Methodological, 18*(4), 339-355. doi:https://doi.org/10.1016/0191-2615(84)90016-X

CDC. (2020). Guidance for Child Care Programs that Remain Open. Retrieved from https://www.cdc.gov/coronavirus/2019-ncov/community/schools-childcare/guidance-for-childcare.html

Christensen, K. E., & Staines, G. L. (1990). Flextime:A Viable Solution to Work/Family Conflict? *Journal of Family Issues, 11*(4), 455-476. doi:10.1177/019251390011004007

de Palma, A., & Lindsey, R. (2002). Comparison of Morning and Evening Commutes in the Vickrey Bottleneck Model. *Transportation Research Record: Journal of the Transportation Research Board, 1807*, 26-33. doi:10.3141/1807-04

ECDA. (2020). Keeping Our Children in Preschools Safe Against the 2019 Novel Coronavirus (2019-nCoV). Retrieved from https://www.ecda.gov.sg/PressReleases/Pages/Advisory-to-Parents-Keeping-Our-Children-in-Preschools-Safe-Against-the-2019-Novel-Coronavirus-(2019-nCoV).aspx

Ezra, M., & Deckman, M. (1996). Balancing Work and Family Responsibilities: Flextime and Child Care in the Federal Government. *Public Administration Review, 56*(2), 174-179. doi:10.2307/977205

Fosgerau, M., & Small, K. (2017). Endogenous Scheduling Preferences and Congestion. *International Economic Review, 58*(2), 585-615. doi:http://onlinelibrary.wiley.com/journal/10.1111/%28ISSN%291468-2354/issues

Gainey, T. W., & Clenney, B. F. (2006). Flextime and Telecommuting: Examining Individual Perceptions. *Southern Business Review, 32*(1), 13-21.

Henderson, J. V. (1981). The economics of staggered work hours. *Journal of Urban Economics, 9*(3), 349-364. doi:http://dx.doi.org/10.1016/0094-1190(81)90032-2

Hurdle, V., Hauser, E., & Fargier, P.-h. (1983). Effects of the choice of departure time on road traffic congestion. Theoretical approach. *Transportation and traffic theory, 8*, 223-263.

Hurdle, V. F. (1981). Equilibrium Flows on Urban Freeways. *Transportation Science, 15*(3), 255-293. doi:10.1287/trsc.15.3.255

Jia, Z., Wang, D. Z. W., & Cai, X. (2016). Traffic managements for household travels in congested morning commute. *Transportation Research Part E: Logistics and Transportation Review, 91*, 173-189. doi:10.1016/j.tre.2016.04.005

Karin, A., & Danielle, H. (2007). *Flexible Work Arrangements in Asia*. Retrieved from https://www.bc.edu/content/dam/files/centers/cwf/research/publications3/researchreports/Flexible%20Work%20Arrangements%20in%20Asia

Li, Z.-C., Lam, W. H. K., & Wong, S. C. (2014). Bottleneck model revisited: An activity-based perspective. *Transportation Research Part B: Methodological, 68*, 262-287. doi:10.1016/j.trb.2014.06.013

Liu, W., Zhang, F., & Yang, H. (2016). Modeling and managing morning commute with both household and individual travels. *Transportation Research Part B: Methodological*. doi:10.1016/j.trb.2016.12.002

Lucas, J. L., & Heady, R. B. (2002). Flextime Commuters and Their Driver Stress, Feelings of Time Urgency, and Commute Satisfaction. *Journal of Business and Psychology, 16*(4), 565-571. doi:10.1023/a:1015402302281

MOM. (2020). *Singapore yearbook of manpower statistics*. (0129-2420). Retrieved from https://stats.mom.gov.sg/Pages/Singapore-Yearbook-Of-Manpower-Statistics-2020.aspx.

Mun, S.-i., & Yonekawa, M. (2006). Flextime, Traffic Congestion and Urban Productivity. *Journal of Transport Economics and Policy, 40*(3), 329-358.

Ralston, D. A. (1989). The benefits of Flextime: Real or imagined? *Journal of Organizational Behavior, 10*(4), 369-373. doi:doi:10.1002/job.4030100407

Small, K. A. (1982). The scheduling of consumer activities: work trips. *The American Economic Review, 72*(3), 467-479.

Su, Q., & Wang, D. Z. W. (2020). On the commute travel pattern with compressed work schedule. *Transportation Research Part A: Policy and Practice, 136*, 334-356. doi:https://doi.org/10.1016/j.tra.2020.04.014

Takayama, Y. (2015). Bottleneck congestion and distribution of work start times: The economics of staggered work hours revisited. *Transportation Research Part B: Methodological, 81, Part 3*, 830-847. doi:http://dx.doi.org/10.1016/j.trb.2015.07.021

Vickrey, W. (1973). *Pricing, metering, and efficiently using urban transportation facilities*. https://trid.trb.org/view/92531

Vickrey, W. S. (1969). Congestion Theory and Transport Investment. *The American Economic Review, 59*(2), 251-260.

Wilson, P. W. (1992). Residential location and scheduling of work hours. *Journal of Urban Economics, 31*(3), 325-336. doi:http://dx.doi.org/10.1016/0094-1190(92)90060-X

Xiao, L., Liu, R., & Huang, H. (2014). Congestion behavior under uncertainty on morning commute with preferred arrival time interval. *Discrete Dynamics in Nature and Society, 2014*.

Zhang, F., Liu, W., Wang, X., & Yang, H. (2017). A new look at the morning commute with household shared-ride: How does school location play a role? *Transportation Research Part E: Logistics and Transportation Review, 103*, 198-217. doi:https://doi.org/10.1016/j.tre.2017.05.004

Zhu, T., Long, J., & Liu, H. (2018). Optimal official work start times in activity-based bottleneck models with staggered work hours. *Transportmetrica B: Transport Dynamics*, 1-27. doi:10.1080/21680566.2018.1460881